\documentclass[runningheads]{llncs}
\usepackage{graphicx}

\usepackage{amsmath}
\usepackage{amssymb}
\usepackage{array}
\newcolumntype{H}{>{\setbox0=\hbox\bgroup}c<{\egroup}@{}}

\usepackage{pgfplots}
\definecolor{cbmagenta}{RGB}{31,120,180}
\definecolor{cbgreen}{RGB}{178,223,138}
\usepgfplotslibrary{units}
\pgfplotsset{
compat=newest
}

\def\imwidth{0.105}

\usepackage{multirow}
\usepackage{hyperref}
\usepackage[tophrase={~to~},binary-units=true,multi-part-units=single]{siunitx}
\usepackage{xcolor}
\begin{document}
\title{Simultaneous Estimation of X-ray Back-Scatter and Forward-Scatter using Multi-Task Learning}
\titlerunning{Multi-Task Scatter Estimation}
%
\author{
     Philipp~Roser\inst{1,4}\and
     Xia~Zhong\inst{2}\and
     Annette~Birkhold\inst{3}\and
     Alexander~Preuhs\inst{1}\and
     Christopher~Syben\inst{1}\and
     Elisabeth~Hoppe\inst{1}\and
     Norbert~Strobel\inst{5}\and
     Markus~Kowarschik\inst{3}\and
     Rebecca~Fahrig\inst{3}\and
     Andreas~Maier\inst{1,4}
}
%
\authorrunning{Philipp Roser et al.}
%
 \institute{
     Pattern Recognition Lab, Friedrich-Alexander-Universit\"at Erlangen-N\"urnberg (FAU), Germany \\ 
        \email{philipp.roser@fau.de}\and
     Diagnostic Imaging, Siemens Healthcare GmbH, Erlangen, Germany \and
     Advanced Therapies, Siemens Healthcare GmbH, Forchheim, Germany \and
     Erlangen Graduate School in Advanced Optical Technologies (SAOT), Germany \and
     Institute of Medical Engineering, University of Applied Sciences Würzburg‐Schweinfurt, Germany
 }
 
\maketitle              
\begin{abstract}
Scattered radiation is a major concern impacting X-ray image-guided procedures in two ways.
First, back-scatter significantly contributes to patient (skin) dose during complicated interventions.
Second, forward-scattered radiation reduces contrast in projection images and introduces artifacts in \mbox{3-D} reconstructions.
While conventionally employed anti-scatter grids improve image quality by blocking X-rays, the additional attenuation due to the anti-scatter grid at the detector needs to be compensated for by a higher patient entrance dose. 
This also increases the room dose affecting the staff caring for the patient. 
For skin dose quantification, back-scatter is usually accounted for by applying pre-determined scalar back-scatter factors or linear point spread functions to a primary kerma forward projection onto a patient surface point.
However, as patients come in different shapes, the generalization of conventional methods is limited.
Here, we propose a novel approach combining conventional techniques with learning-based methods to simultaneously estimate the forward-scatter reaching the detector as well as the back-scatter affecting the patient skin dose. 
Knowing the forward-scatter, we can correct X-ray projections, while a good estimate of the back-scatter component facilitates an improved skin dose assessment.
To simultaneously estimate forward-scatter as well as back-scatter, we propose a multi-task approach for joint back- and forward-scatter estimation by combining X-ray physics with neural networks.
We show that, in theory, highly accurate scatter estimation in both cases is possible. 
In addition, we identify research directions for our multi-task framework and learning-based scatter estimation in general.
\keywords{X-ray scatter \and Skin dose \and Multi-task learning}
\end{abstract}
\section{Introduction}
\label{sec:introduction}
X-ray fluoroscopic guidance enables minimally-invasive interventions. Unfortunately, photons scattered by the patient impair X-ray image quality and increase the X-ray dose affecting both patient as well as staff. There are two major types of scatter in X-ray imaging: back-scatter and forward-scatter.

Back-scatter contributes up to \SIrange{30}{60}{\percent} of the total skin dose \cite{Henss:1998:BSF}.
Unless properly accounted for, it impairs accurate online monitoring of skin dose, which is a critical means in dose management for interventional fluoroscopic imaging.
By providing constant feedback on accumulated skin dose values to the physicians, they can spread the dose using table movements and C-arm rotations to avoid excessive peak skin dose values.
Therefore, most X-ray imaging systems are equipped with a dose chamber measuring the kinetic energy released per unit mass in air (typically referred to as air kerma). 
These measured values can be used to calibrate either on-site Monte Carlo (MC) simulations \cite{LoyRodas:2015:Scatter}, U-Net-accelerated dose simulations 
\cite{Ronneberger:2015:Unet,Roser:2019:DoseLearning}, or kerma forward projection (KFP) onto a digital patient model \cite{Balter:2006:SkinDose}.
Simulation approaches can yield more accurate dose estimates if a precise model of the actual imaging setting is available.
Unfortunately, prior knowledge, such as the exact patient anatomy, is not available in general.
Therefore, current state-of-the-art systems rely on patient shape models, KFP, and several correction terms accounting for the inverse square law, skin absorption, and back-scatter \cite{Balter:2006:SkinDose,Johnson:2011:SkinDose,Rana:2016:SkinDose}.
Previous studies have shown that these back-scatter factors (BSF), determined experimentally or using MC simulations, have the potential to increase the accuracy of skin dose estimation \cite{Henss:1998:BSF,Rana:2016:SkinDose}. 
Yet, BSFs are highly dependent on the imaging setting and specific patient. 
Despite this fact, they are usually pre-computed using MC simulation or measured empirically yielding large tables of BSFs to cover the whole patient population, anatomic regions, and X-ray characteristics.
These tables are cumbersome to obtain and maintain.
Furthermore, the rich source of information reflecting patient as well as X-ray beam characteristics contained in the X-ray images themselves is not used in any form by these dose estimation methods.

Forward-scatter, on the other hand, causes uneven exposure and loss of contrast in X-ray images.
This is why, hardware- and/or software-based scatter correction methods are used to enhance image quality \cite{Ruehrnschopf:2011:ScatterCompensation2,Ruehrnschopf:2011:ScatterCompensation1}.
Typically, anti-scatter grids are used to physically block scattered X-rays.
However, since they also absorb some primary radiation, a higher patient entrance dose is needed to maintain the desired X-ray exposure level at the detector \cite{Chan:1982:ASG}.
With an increased focus on X-ray dose reduction, protection, and risk management, grid-less X-ray imaging is desirable.
In particular, in pediatrics, where patient X-ray dose plays a crucial role and where patients are usually smaller, anti-scatter grids are commonly removed \cite{Fritz:2014:ASGPed,Ubeda:2013:ASGPed}.
Software-only approaches, such as the recently introduced single-task U-Net-based deep scatter estimation (DSE)  \cite{Maier:2019:DSE}, might render conventional approaches \cite{Li:2008:KSE,Ohnesorge:1999:KSE,Sun:2010:KSE} obsolete.

In this work, we apply convolutional neural networks (CNN) to estimate back-scatter as well as forward-scatter.
To this end, we propose a multi-task framework to calculate back- and forward-scatter in a one-step procedure by combining X-ray physics' models with modern learning-based methods.
Since back- and forward-scatter share the same mathematical description, we can leverage learning-based forward-scatter estimation to also infer back-scatter directly from an X-ray projection image and a patient shape model.

\subsubsection{Contribution}
To the best of our knowledge, this paper presents several novel ideas not yet published elsewhere: (1) learning-based back-scatter estimation using a patient model, (2) deriving back-scatter from an X-ray image projection, and (3) simultaneous back-scatter and forward-scatter estimation.
Finally, we propose a lightweight network architecture to efficiently implement it reaching an accuracy comparable to outcomes obtained using a multi-task U-Net.

\section{Material and Methods}
\label{sec:methods}
\begin{figure}[tb]
    \centering
    \includegraphics[clip,trim=0 0 1cm 0,width=\textwidth]{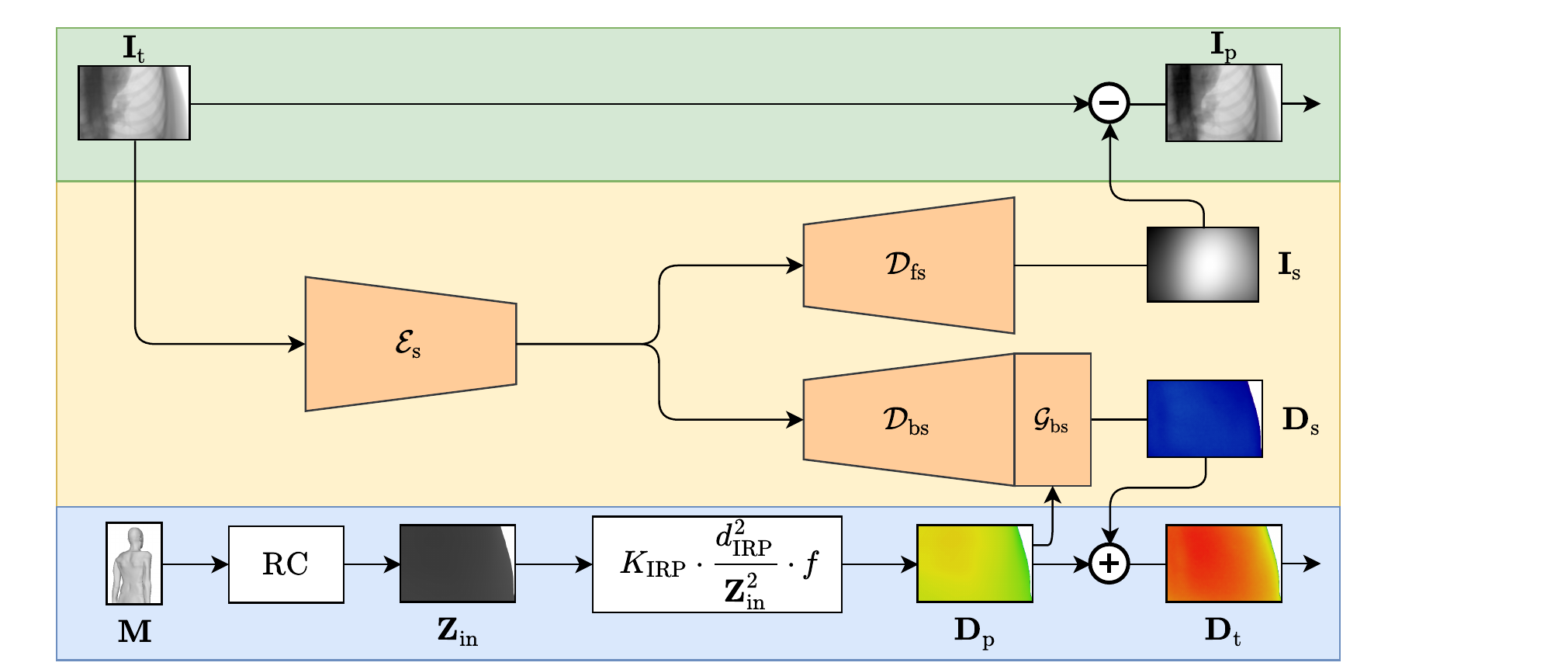}
    \caption{
    Overview of the proposed method with a dedicated dual autoencoder-like architecture (DAE). 
    Based on the patient shape model and the imaging geometry of the X-ray system, the ray-caster (RC) estimates an entrance map $\vec{Z}_\text{in}$. 
    Based on $\vec{Z}_\text{in}$, the primary skin dose $\vec{D}_\text{p}$ is calculated using conventional KFP.
    An encoder $\mathcal{E}_\text{s}$ extracts the latent representation of the underlying scatter distribution. 
    Using two independent decoders $\mathcal{D}_\text{fs}$ and $\mathcal{D}_\text{bs}$ forward-scatter $\vec{I}_\text{s}$ and back-scatter $\vec{D}_\text{s}$ are estimated, respectively.
    To account for the domain transfer to dose, $\mathcal{D}_\text{bs}$ is extended by an additional convolutional block $\mathcal{G}_\text{bs}$ with $\vec{D}_\text{p}$ as second input.
    Knowing the $\vec{I}_\text{s}$, we can estimate the primary image $\vec{I}_\text{p}$, while $\vec{D}_\text{s}$ can be used to calculate the total skin dose $\vec{D}_\text{t}$.
    }
    \label{fig:meth:outline}
\end{figure}

Figure \ref{fig:meth:outline} depicts the outline of the proposed method.
The basic principle is to exploit our rich understanding of the photon interactions causing X-ray forward-scatter and back-scatter, respectively.
Since both are caused by same underlying scattering interactions, it is reasonable to estimate their effects on image formation and skin dose together.
Unfortunately, analytic and stochastic scatter estimation using established physics models is time-consuming and relies on accurate prior knowledge on the patient anatomy \cite{Baer:2012:HSE,Poludniowski:2009:MCScatter,Wang:2018:Acuros}.
This can, however, be done faster and with a comparable accuracy, if we combine the underlying physics with a data-driven multi-task CNN.
In particular, our approach involves KFP and a multi-task scattering model. 
Details are described below.

\subsubsection{Kerma Forward Projection}
\label{sec:kfp}
As in conventional dose-monitoring systems, a \mbox{3-D} patient shape model $\mathcal{M}$ is at the heart of our method.
Potential sources of such a digital twin are a pre-operative computed tomography (CT) scan, a point cloud reconstruction based on \mbox{3-D}-capable camera systems, or an active shape model based on meta-parameters such as weight, height, and age \cite{Johnson:2011:SkinDose,LoyRodas:2015:Scatter,Zhong:2017:Model}.
In the following, we assume that we are given an already registered patient shape model.
Based on the patient shape model and X-ray system geometry, we calculate the distance per detector pixel, where the X-ray enters the patient $\vec{Z}_\text{in} \in \mathbb{R}^{w\times h}$, with the width $w$ and height $h$ of the detector. 
To make our method robust and flexible in a clinical situation, we use an efficient grid traversal algorithm \cite{Amanatides:1987:RC}, which allows for patient shapes defined by either a mesh, implicit or explicit analytical functions, or tomographic data.
For each detector pixel, a ray iteratively traverses a cubic grid from the X-ray source position to the respective pixel.
Once the ray intersects the patient model, the traveled (source-patient-surface) distance is stored in the patient entrance map $\vec{Z}_\text{in}$.
Once $\vec{Z}_\text{in}$ is known, the primary component of the skin dose per detector pixel $\vec{D}_\text{p} \in \mathbb{R}^{w\times h}$ (patient entrance dose) is computed by applying the inverse square law \cite{Balter:2006:SkinDose}.
The air-kerma measured at the interventional reference point (IRP) $K_\text{IRP}$ in \si{\milli\gray} or \si{\joule\per\gram} is projected onto the skin surface yielding
\begin{equation}
    \vec{D}_\text{p} = K_\text{IRP} \cdot d_\text{IRP}^2/\vec{Z}_\text{in}^2 \cdot f,
\end{equation}
where $d_\text{IRP}$ is the distance between the X-ray source and the IRP and $f$ is a unit-less tissue-conversion factor which can be pre-calculated for any X-ray spectrum.
Unfortunately, for back-scatter $\vec{D}_\text{s}$, no such analytical model exists without extensive prior knowledge on the patient anatomy.
However, we can make use of the rich information encoded in the measured X-ray projection image.

\subsubsection{Multi-Task Scattering Model}
\begin{figure}
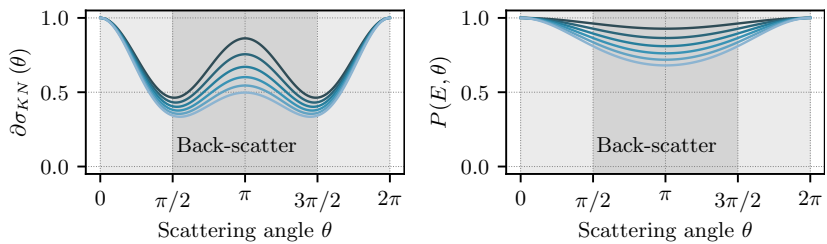

    \centering
    \begin{tabular}{cc}
        \resizebox{0.45\textwidth}{!}{
            \input{KN.pgf}
        } 
        & 
        \resizebox{0.45\textwidth}{!}{
            \input{E.pgf}
        } \\
    \end{tabular}
    \caption{Normalized KN CS $\partial\sigma_\text{KN}(\theta)$, left, and energy ratio $P(E, \theta)$, right, plotted against the scattering angle $\theta$ for $E \in \{20, 40, \dots, 120\}$ in \si{\kilo\eV} (dark blue to light blue).}
    \label{fig:scattering}
\end{figure}

Since, in the medical X-ray energy regime, the incoherent Compton scatter dominates over the coherent Rayleigh scatter, we can safely neglect the latter in the following considerations.
The occurence probability of scattering interactions can be expressed in terms of cross-sections (CS) $\sigma$.
The Compton scattering model for an infinitesimal volume element $\partial\Omega$ is based on the differential Klein-Nishina (KN) CS $\frac{\partial\sigma_\text{KN}}{\partial\Omega}$, given by
\begin{equation}
    \frac{\partial\sigma_\text{KN}}{\partial\Omega} = 0.5\,r^2_e\,P(E, \theta)^2 \left[P(E, \theta) + P(E, \theta)^{-1} - \sin^2 \theta \right]
    \enspace ,
\label{eq:KN}
\end{equation}
with the classical electron radius $r_e$, 
the scattering angle $\theta$, and the ratio of photon energy $E$ after and before the interaction $P(E, \theta)$.
The ratio $P(E, \theta)$ is defined as
\begin{equation}
    P(E,\theta) = \frac{1}{1 + (\frac{E}{m_ec^2})(1-\cos\theta)} \enspace ,
    \label{eq:ratio}
\end{equation}
where $m_e$ denotes the electron rest mass and $c$ the speed of light, respectively.
As Fig. \ref{fig:scattering} shows, the KN model has several useful properties.

First, since Eq. \ref{eq:KN} only depends on energy ratios and symmetric trigonometric functions, we can establish a functional relationship between back-scatter and forward-scatter.
In other words, given the forward-scatter distribution, the back-scatter distribution is for infinitesimal volumes analytically described by the KN formula.
From recent studies \cite{Maier:2019:DSE}, we know that forward-scatter $\vec{I}_\text{s} = \mathcal{U}_\text{ST}(\vec{I}_\text{t})$ in \si{\joule\per\cm\squared} can be extracted from the measured X-ray projection $\vec{I}_\text{t}$ using a single-task U-Net $\mathcal{U}_\text{ST}$.
Hence, we can relate forward- and back-scatter via
\begin{equation}
    \left(\frac{\mu_\text{en}}{\rho}\right)^{-1} \vec{D}_\text{s} \sim \vec{I}_\text{s} = \mathcal{U}_\text{ST}(\vec{I}_\text{t}) \enspace ,
\end{equation}
with the mass energy absorption coefficient $\left(\frac{\mu_\text{en}}{\rho}\right)$ in \si{\cm\squared\per\gram}.
Since $\left(\frac{\mu_\text{en}}{\rho}\right)$ relates to a simple linear scaling, we can omit it below.
Unfortunately, we do not know a simple yet accurate model of this relationship for arbitrary patient anatomies.
However, being based on the same physical effects, it can be concluded that both forward- and back-scatter can be encoded by similar features or latent variables.
Consequently, it appears attractive to learn both in a multi-task fashion.
Since the U-Net has yielded promising results for forward-scatter estimation, we investigate its applicability to the task at hand.
Therefore, by supplying the primary skin dose estimate $\vec{D}_\text{s}$ we extend the U-Net to an multi-task function $(\vec{D}_\text{s}, \vec{I}_\text{s}) = \mathcal{U}_\text{MT}(\vec{D}_\text{p}, \vec{I}_\text{t})$.
The U-Net comprises six levels, two convolutional layers per block, average pooling, \num{16} initial feature maps doubled after each pooling operation, rectified linear units (ReLU) \cite{Glorot:2011:ReLU} as activations, and \num{31}\,M parameters to train in total.
Although the U-Net has proven to be a powerful function approximator for numerous tasks, its high parameter complexity and degree of connectivity is not easy to fully comprehend making it a potentially risky tool for clinical image processing.

Especially the U-Net's skip-connections lead to outputs covering the whole frequency spectrum.
However, a Fourier analysis of Eqs. \ref{eq:KN} and \ref{eq:ratio} for the diagnostic X-ray energy regime reveals mostly low-frequency characteristics.
This observation is substantiated by the low amplitude and smoothness of the corresponding plots in Fig. \ref{fig:scattering}.
Therefore, we propose to degenerate the U-Net to a dual autoencoder-like CNN (DAE) without skip-connections to constrain its output frequency, as depicted in Fig. \ref{fig:meth:outline}.
While we keep the encoding path $\mathcal{E}_\text{s}$ to extract the latent scatter distribution from $\vec{I}_\text{t}$, we split the decoding paths $\mathcal{D}_\text{bs}$ and $\mathcal{D}_\text{fs}$ to separately estimate $\vec{D}_\text{s}$ and $\vec{I}_\text{s}$, respectively.
Since forward- and back-scatter are based on the same particle interactions, it is reasonable for both to share the same latent space, while the decoders can be interpreted as opposing projections on either the detector (forward-scatter) or the patient skin surface (back-scatter). 
To further enforce low-frequency characteristics via a compact latent space, the number of feature maps is not doubled per down-sampling operation as it is typically done for the U-Net.
Similar to $\mathcal{U}_\text{MT}$, the encoder $\mathcal{E}_\text{s}$ consists of six convolutional blocks (two layers with \num{16} \num{3x3} kernels and ReLU activations) with average pooling operations in between.
The decoders resemble $\mathcal{E}_\text{s}$ with bilinear up-sampling instead of average pooling. 
In addition, we extend $\mathcal{D}_\text{bs}$ by an additional convolutional block $\mathcal{G}_\text{bs}$ with $\vec{D}_\text{p}$ as second input to account for the domain transfer to skin dose.
The number of parameters to train is with \num{105}\,k two magnitudes lower than the U-Net's. 
In general, $\vec{D}_\text{s}$ and $\vec{I}_\text{s}$ have both different co-domains and SI units. 
To circumvent error-prone loss weighting, we used the mean absolute percentage error (MAPE) cost function.

\section{Results and Discussion}
\subsubsection{Data}
As it is technically not feasible to extract matching pairs of X-ray projections in a clinical setup, we based our experiments on synthetic data.
To this end, we simulated data using a MC photon transport code yielding X-ray images, Compton and Rayleigh scatter, and \mbox{3-D} kerma distributions \cite{Badal:2009:MCGPU,Roser:20:MCPreProcessing}.
As patient models, we selected 22 head scans from the HNSCC-3DCT-RT data set \cite{Bejanero:18:Heads} and 17 thorax scans from the CT-Lymph-Nodes data set \cite{Roth:15:Lymphs}, both provided by The Cancer Imaging Archive (TCIA) \cite{Clark:13:TCIA}.
We used a \SI{100}{\kilo\volt}p spectrum and simulated \num{5e10} primary photons for each X-ray image.
All images comprise \num{256 x 384} pixels with an area of \SI{1.16 x 1.03}{\mm\squared}.
For each patient, we computed \num{60} projections with three source positions, \num{20} projection angles (\SIrange{0}{95}{\degree}, \SI{5}{\degree} sampling), and fixed source-to-detector distance (\SI{130}{\cm}).
In total, \num{2340} data points were available.
We separated one patient for validation and two patients for testing for both data sets.

\subsubsection{Experimental Setup}
To thoroughly assess our results and provide for an ablation study, we compared our method to several baseline networks: (a) a lean and straightforward single-task autoencoder-like network similar to one path of our DAE network (AE, six levels, two convolutional layers per block, average pooling, \num{16} feature maps per convolution, ReLU activation, \num{49}\,k parameters to train) and (b) a single-task U-Net (six levels, two convolutional layers per block, average pooling, \num{16} initial feature maps doubled after each pooling operation, ReLU activation, \num{31}\,M parameters to train).
As inputs, we either considered the measured X-ray projection $\vec{I}_\text{t}$ or the primary dose distribution $\vec{D}_\text{p}$.
As outputs, we either considered the X-ray forward-scatter $\vec{I}_\text{s}$ or the back-scatter skin dose $\vec{D}_\text{s}$, respectively.
In the training phase, we fixed the hyper-parameters for all networks, including our multi-task learning approach.
Therefore, we also minimized the MAPE for both single-task networks.
To optimize the network weights, we used adaptive moments \cite{Kingma:2014:Adam} with a learning rate of \num{e-4} and a batch size of four.
Since this is the first time where back-scatter is estimated in a data-driven manner, we also provide error metrics for BSFs.
To highlight the minimum error bounds achievable using conventional back-scatter correction, we calculated the BSFs specifically for each X-ray projection, which compares to an over-fitting scenario.

\subsubsection{Results}
\begin{table}[tb]
    \centering
    \caption{
    Expected error rates $\mu_\varepsilon(\sigma_\varepsilon)$ for all network settings. 
    The AE and the U-Net are trained to either infer $\vec{I}_\text{s}$ or $\vec{D}_\text{s}$. 
    For dose estimation, $\mu_\varepsilon(\sigma_\varepsilon)$ we also compare to results obtained using back-scatter factors (BSFs) which were optimized for the specific settings. 
    }
    \label{tab:results}
    \begin{tabular}{@{\extracolsep{4pt}} c l c H c c c c H @{}}
        \cline{1-8}
        \cline{1-8}
        & \multirow{2}{*}{Method} & \multirow{2}{*}{Map} & \multirow{2}{*}{Output} & \multicolumn{2}{c}{Head $\mu_\varepsilon(\sigma_\varepsilon)$ [\si{\percent}]} & \multicolumn{2}{c}{Thorax $\mu_\varepsilon(\sigma_\varepsilon)$ [\si{\percent}]} & \multirow{2}{*}{$\mu_t$ [\si{\milli\second}]} \\
        \cline{5-6}\cline{7-8}
        & & & & $\vec{D}_\text{s}$ & $\vec{I}_\text{s}$ & $\vec{D}_\text{s}$ & $\vec{I}_\text{s}$ \\
        \cline{1-8}
        \multirow{5}{*}{\rotatebox{90}{\shortstack{Baseline \\ (single)}}} 
        & BSF 
            & $\vec{D}_\text{p}\mapsto \vec{D}_\text{s}$ & $\vec{D}_\text{s}$ & \num{15.57 \pm 2.17} & - & \num{8.04 \pm 2.08} & - \\
        & AE 
            & $\vec{I}_\text{t}\mapsto \vec{I}_\text{s}$ & $\vec{I}_\text{s}$ & - & \num{5.61 \pm 1.64} & - & \num{10.12 \pm 4.01} \\
            &  & $\vec{D}_\text{p} \mapsto \vec{D}_\text{s}$ & $\vec{D}_\text{s}$ & \num{24.56 \pm 3.80} & - & \num{12.66 \pm 6.20} & - \\
        & U-Net 
            & $\vec{I}_\text{t} \mapsto \vec{I}_\text{s}$ & $\vec{I}_\text{s}$ & - & \num{5.61 \pm 2.55} & - & \num{8.20 \pm 3.71} \\
            & & $\vec{D}_\text{p} \mapsto \vec{D}_\text{s}$ & $\vec{D}_\text{s}$ & \num{11.26 \pm 2.44} & - & \num{9.46 \pm 5.53} & - \\
        \cline{1-8}
        \multirow{3}{*}{\rotatebox{90}{\shortstack{Ours \\ (multi)}}} 
        & AE 
            & $\vec{D}_\text{p},\vec{I}_\text{t} \mapsto \vec{D}_\text{s},\vec{I}_\text{s}$ & $\vec{D}_\text{s},\vec{I}_\text{s}$ & \num{21.98 \pm 4.01} & \num{6.84 \pm 1.59} & \num{10.12 \pm 4.27} & \num{9.46 \pm 2.77} \\
        & U-Net 
            & $\vec{D}_\text{p},\vec{I}_\text{t} \mapsto \vec{D}_\text{s},\vec{I}_\text{s}$ & $\vec{D}_\text{s},\vec{I}_\text{s}$ & \num{9.11 \pm 2.20} & \textbf{4.22(166)} & \textbf{5.22(156)} & \textbf{6.70(280)} \\
        & DAE 
            & $\vec{D}_\text{p},\vec{I}_\text{t} \mapsto \vec{D}_\text{s} , \vec{I}_\text{s}$ & $\vec{D}_\text{s} , \vec{I}_\text{s}$ & \textbf{8.09(142)} & \num{7.87 \pm 2.26} & \num{6.95 \pm 3.33} & \num{9.06 \pm 3.24} \\
        \cline{1-8}
        \cline{1-8}
    \end{tabular}
\end{table}
\begin{figure}[tb]
    \centering
    \begin{tabular}{@{\extracolsep{1pt}} c c c c c c c c c @{}}
    & $\vec{I}_\text{t}$ & $\vec{I}_\text{s}$ & $\hat{\vec{I}}_\text{s}$ & $\vec{\varepsilon}_{\vec{I}}$ & $\vec{D}_\text{p}$ & $\vec{D}_\text{s}$ & $\hat{\vec{D}}_\text{s}$ & $\vec{\varepsilon}_{\vec{D}}$ \\
        \multirow{2}{*}{\rotatebox{90}{\shortstack{U-Net \\ (single)}}} 
            & \includegraphics[width=\imwidth\linewidth]{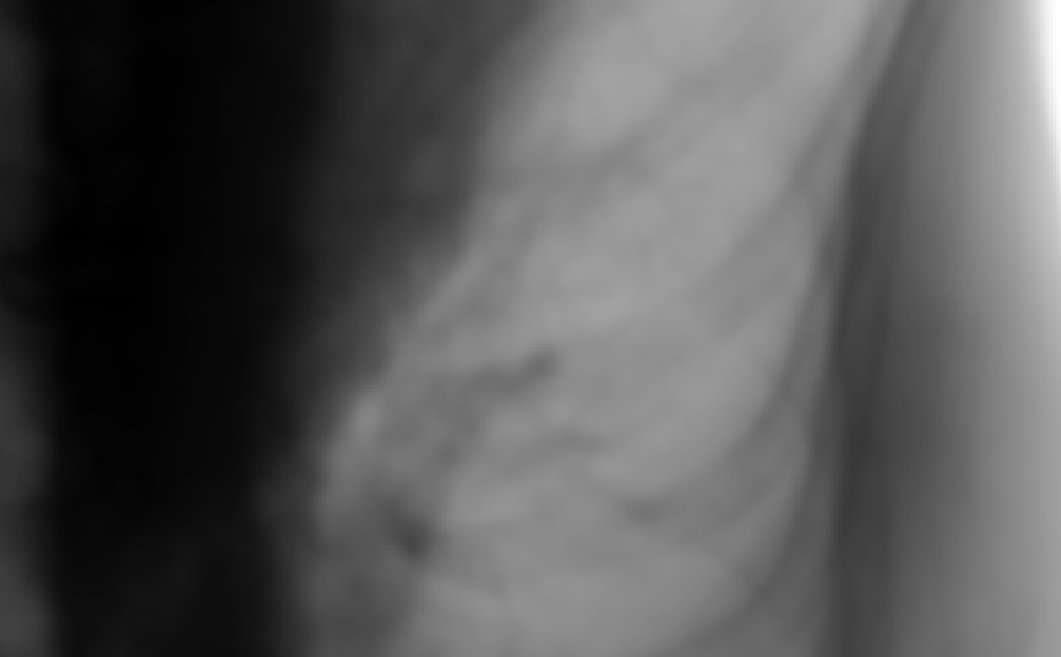}
            & \includegraphics[width=\imwidth\linewidth]{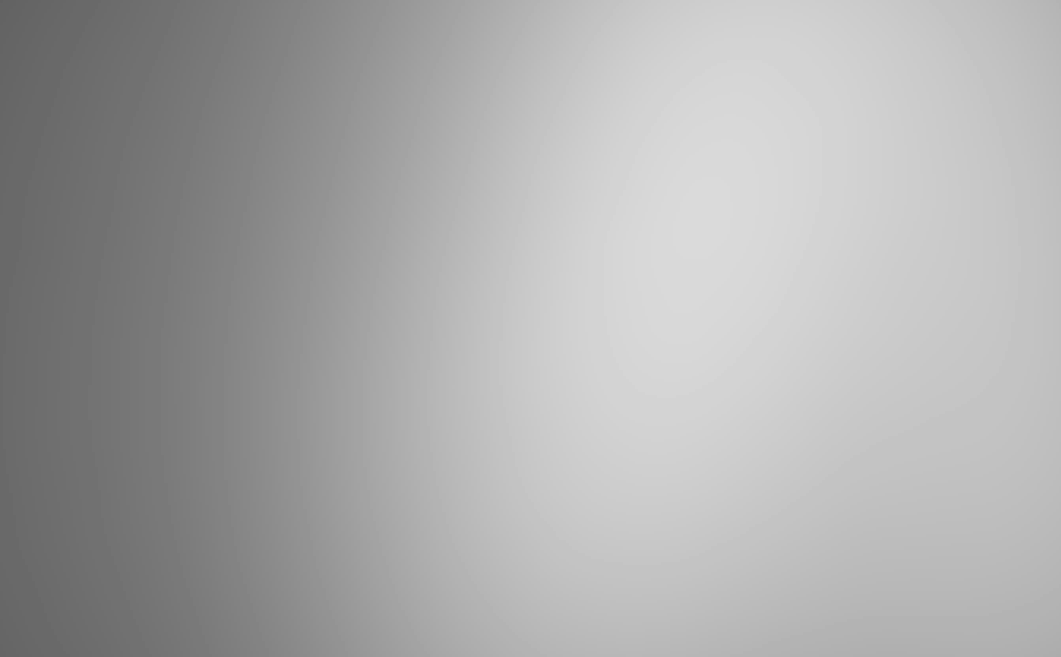}
            & \includegraphics[width=\imwidth\linewidth]{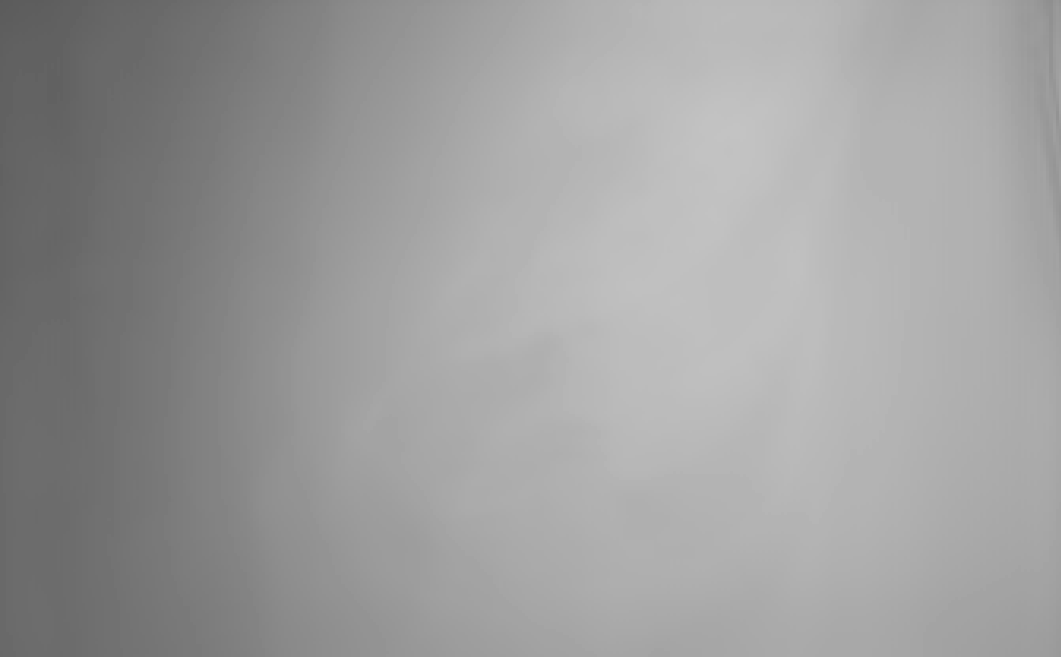}
            & \includegraphics[width=\imwidth\linewidth]{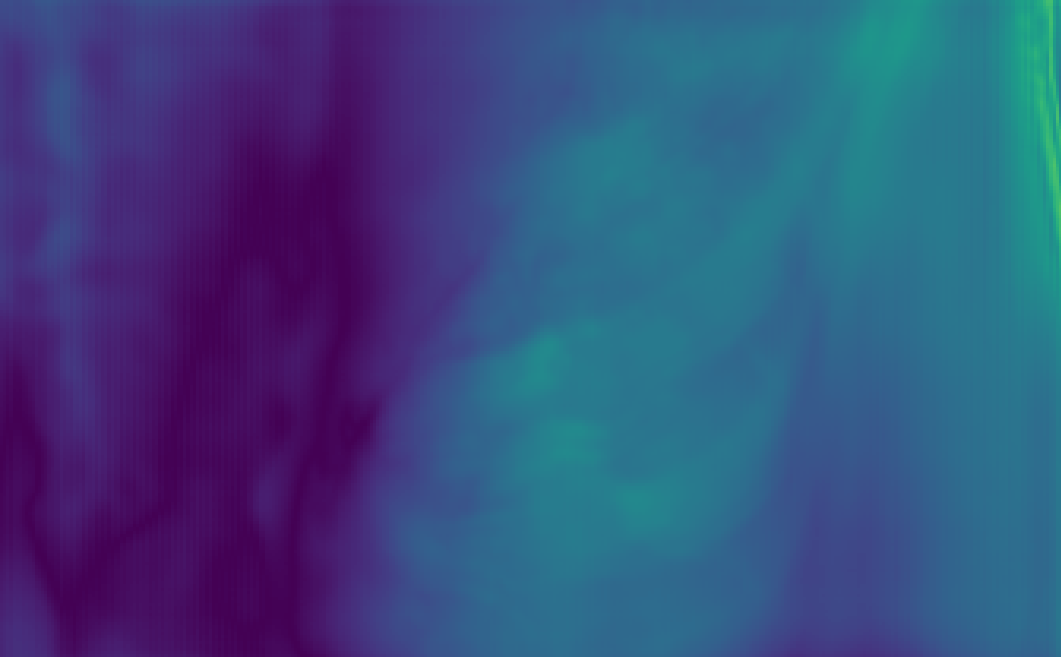}
            & \includegraphics[width=\imwidth\linewidth]{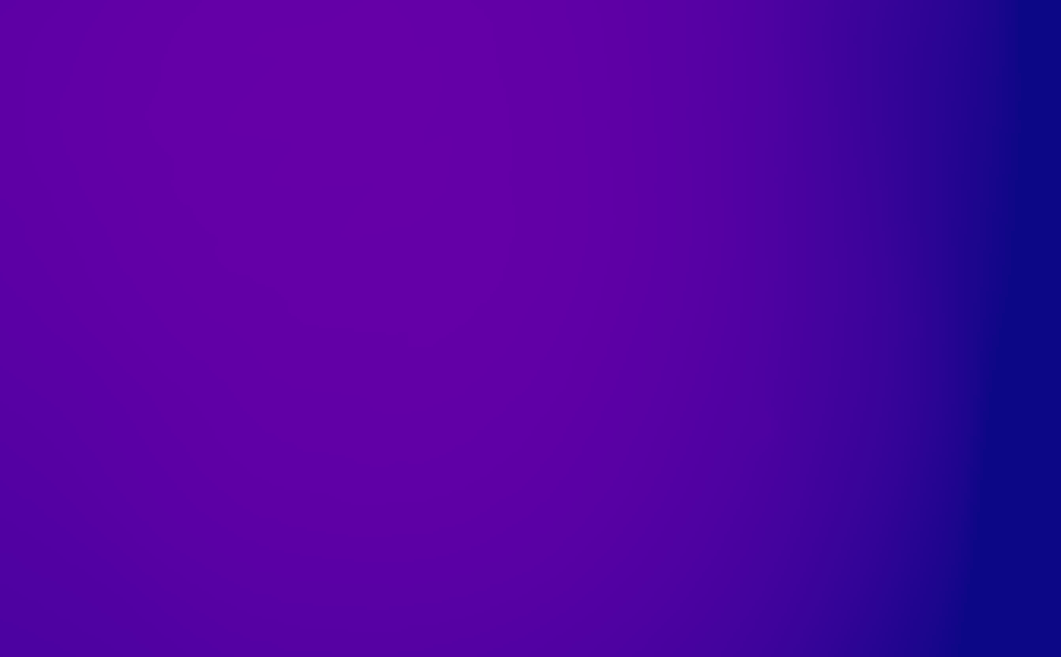}
            & \includegraphics[width=\imwidth\linewidth]{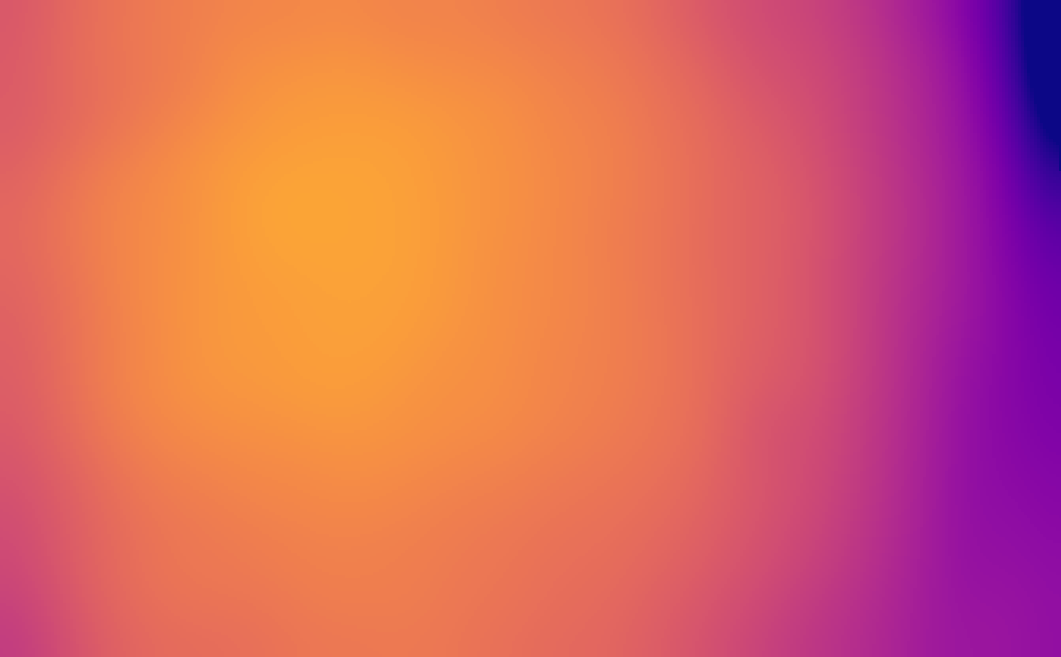}
            & \includegraphics[width=\imwidth\linewidth]{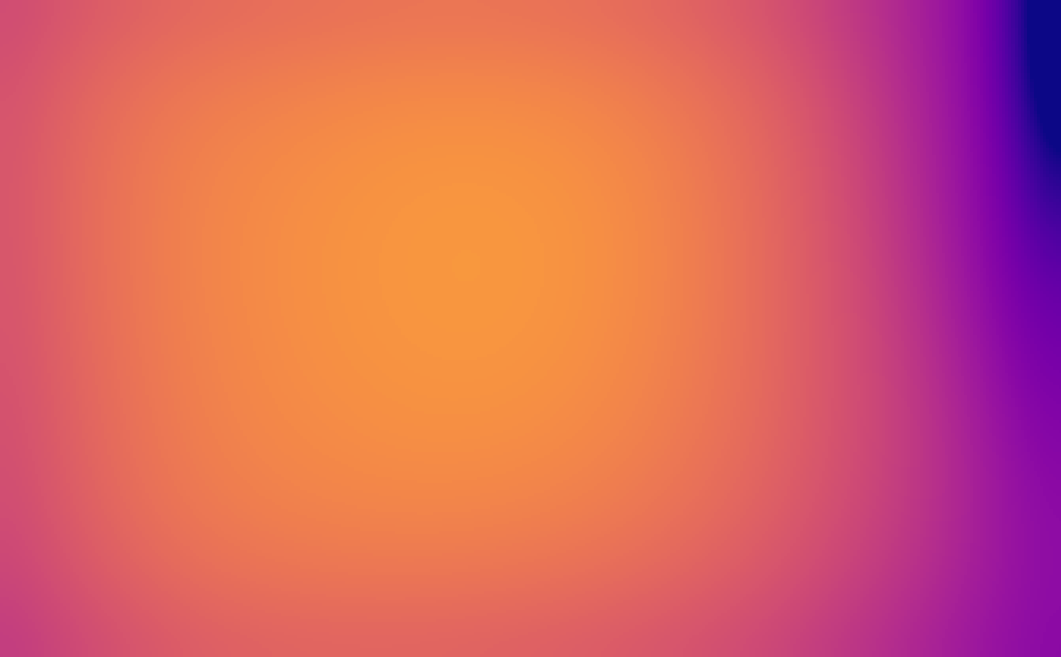}
            & \includegraphics[width=\imwidth\linewidth]{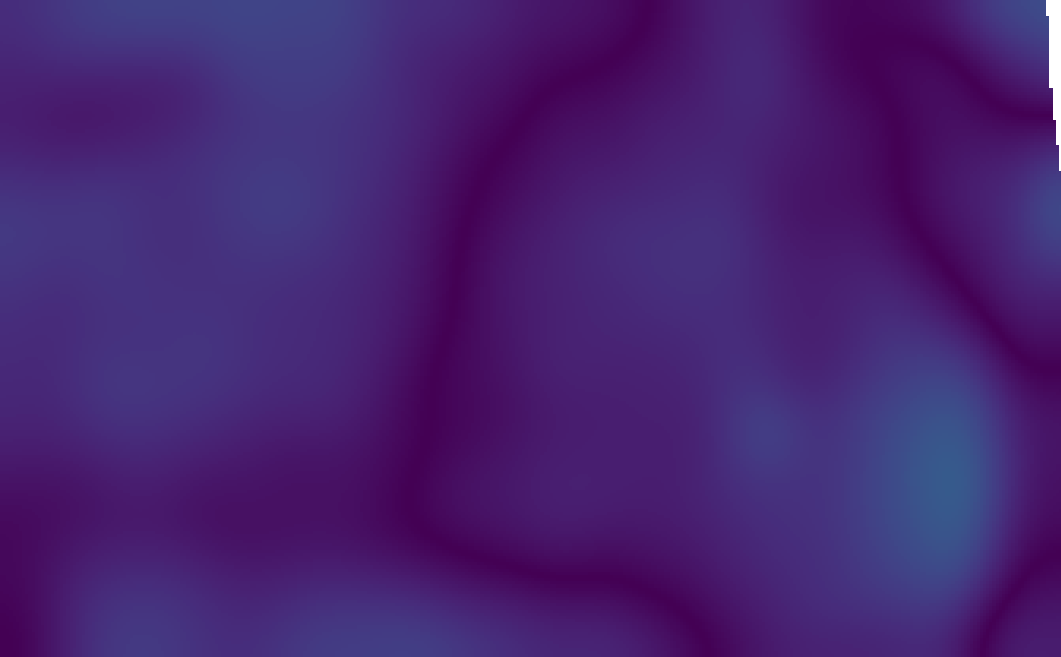} \\
            & \includegraphics[width=\imwidth\linewidth]{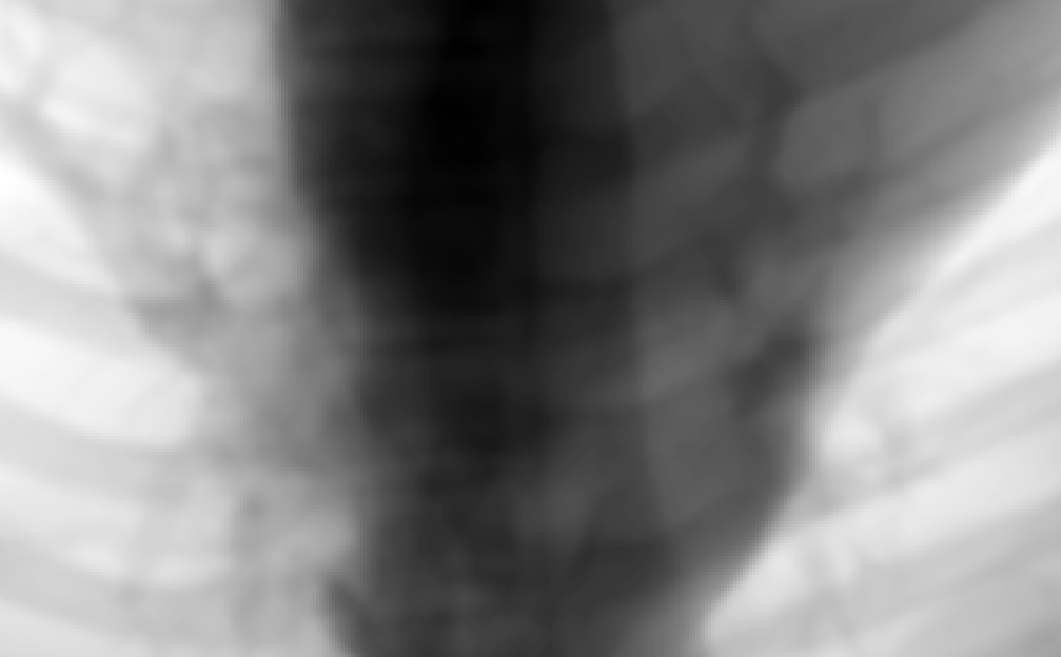}
            & \includegraphics[width=\imwidth\linewidth]{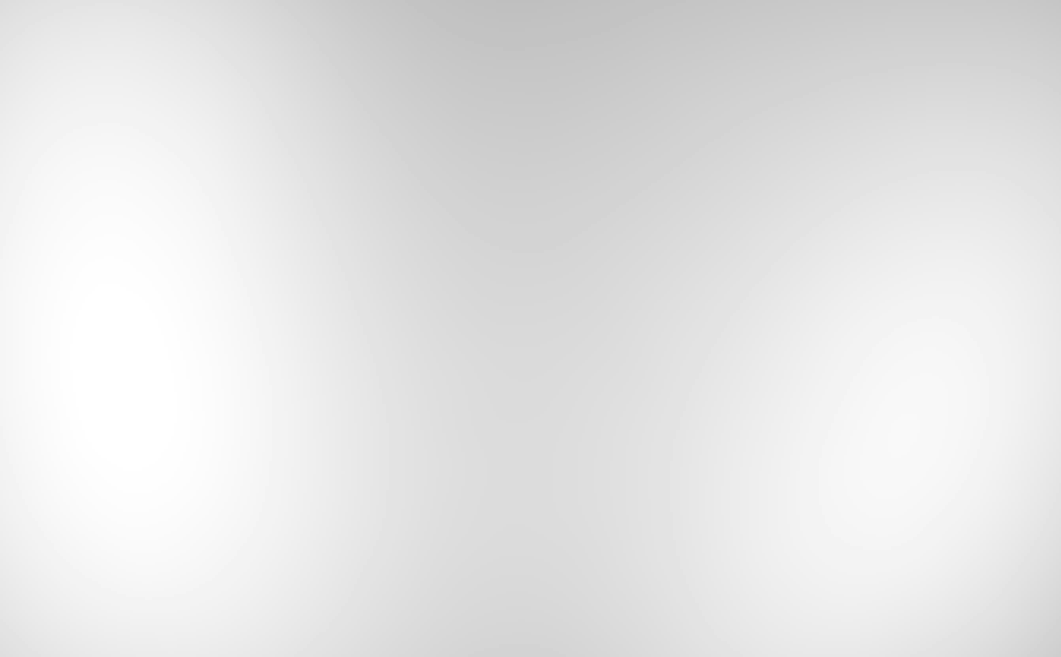}
            & \includegraphics[width=\imwidth\linewidth]{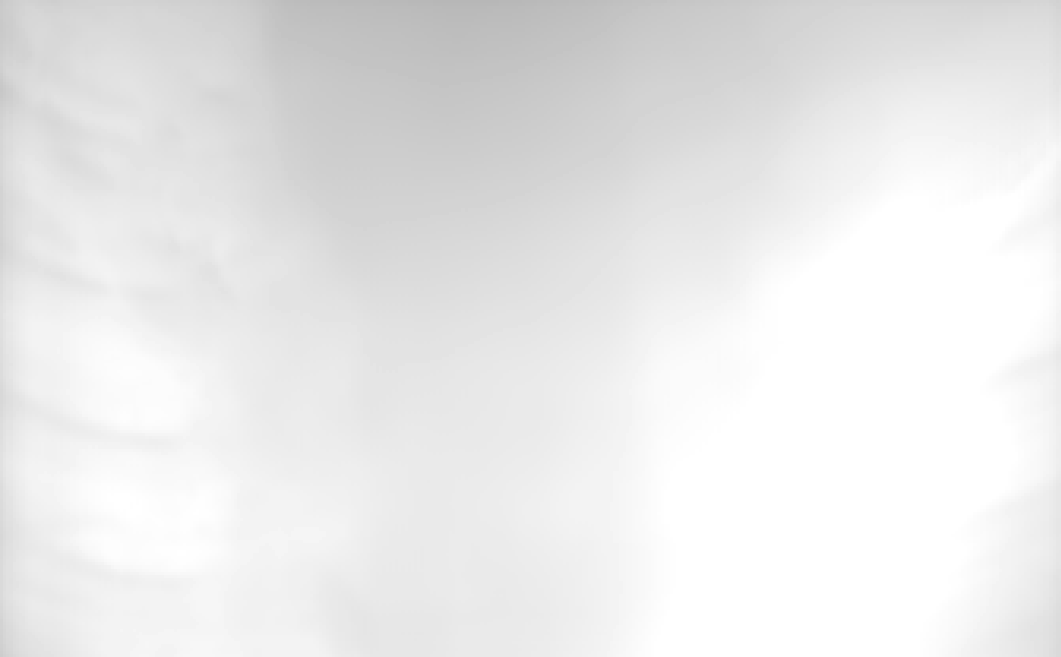}
            & \includegraphics[width=\imwidth\linewidth]{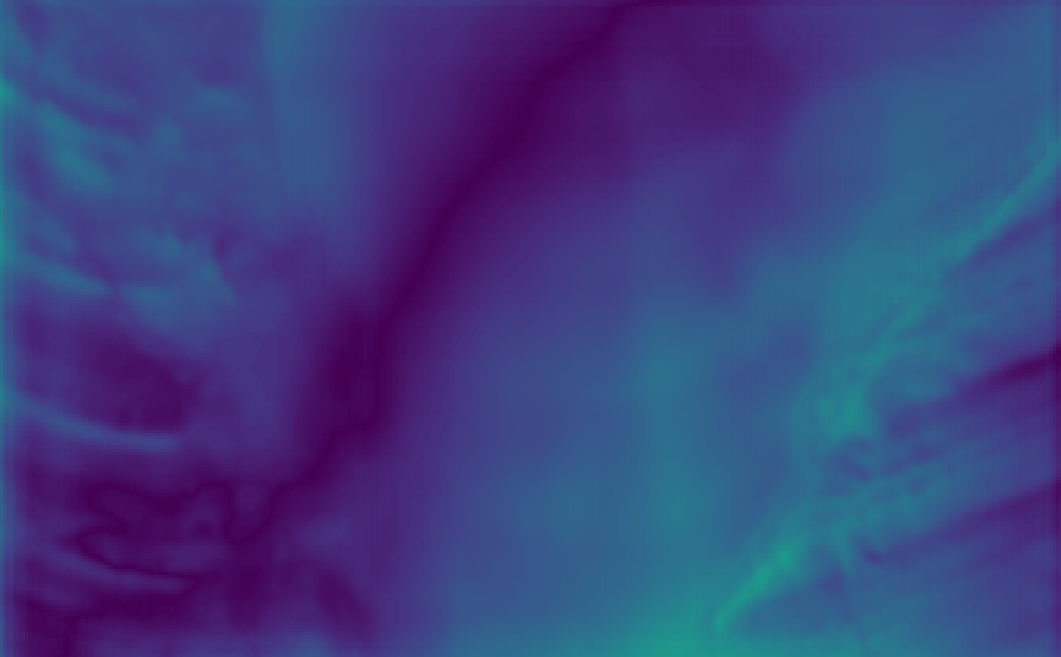}
            & \includegraphics[width=\imwidth\linewidth]{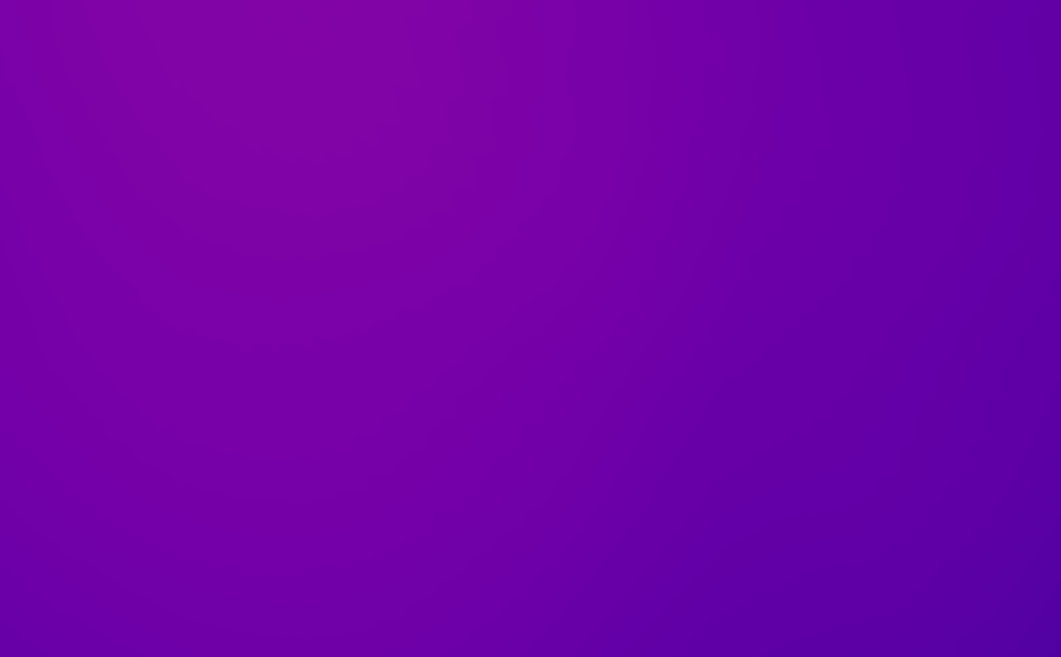}
            & \includegraphics[width=\imwidth\linewidth]{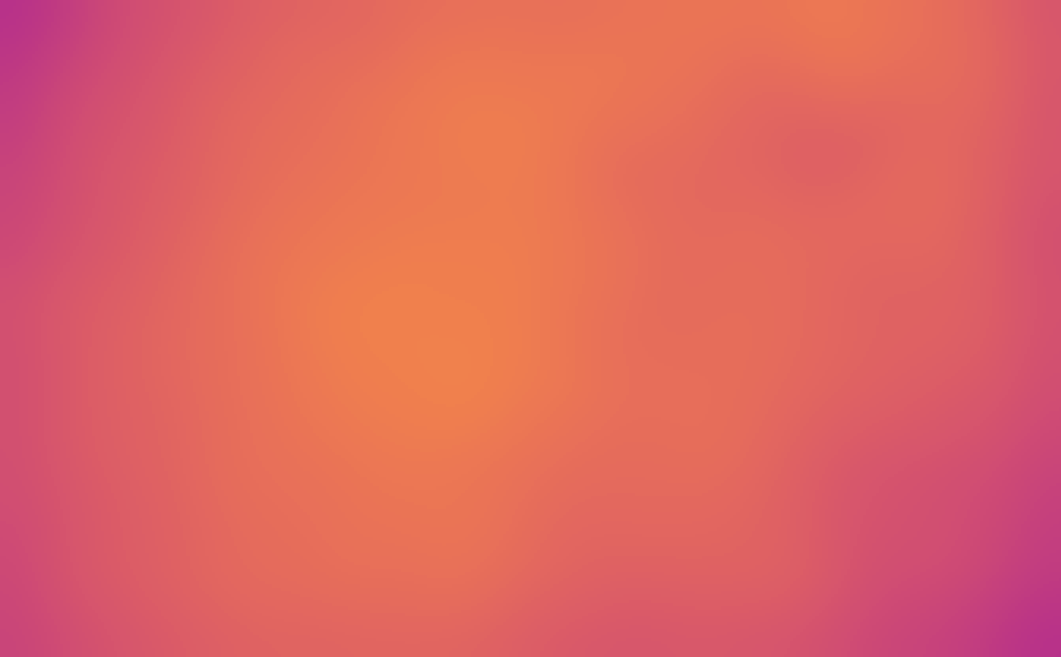}
            & \includegraphics[width=\imwidth\linewidth]{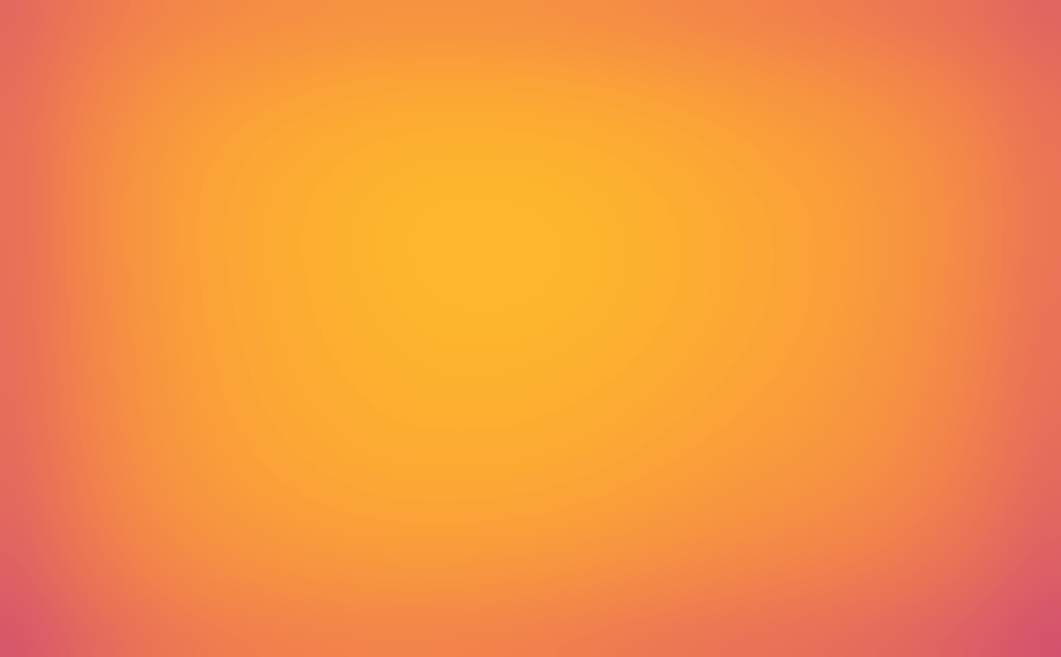}
            & \includegraphics[width=\imwidth\linewidth]{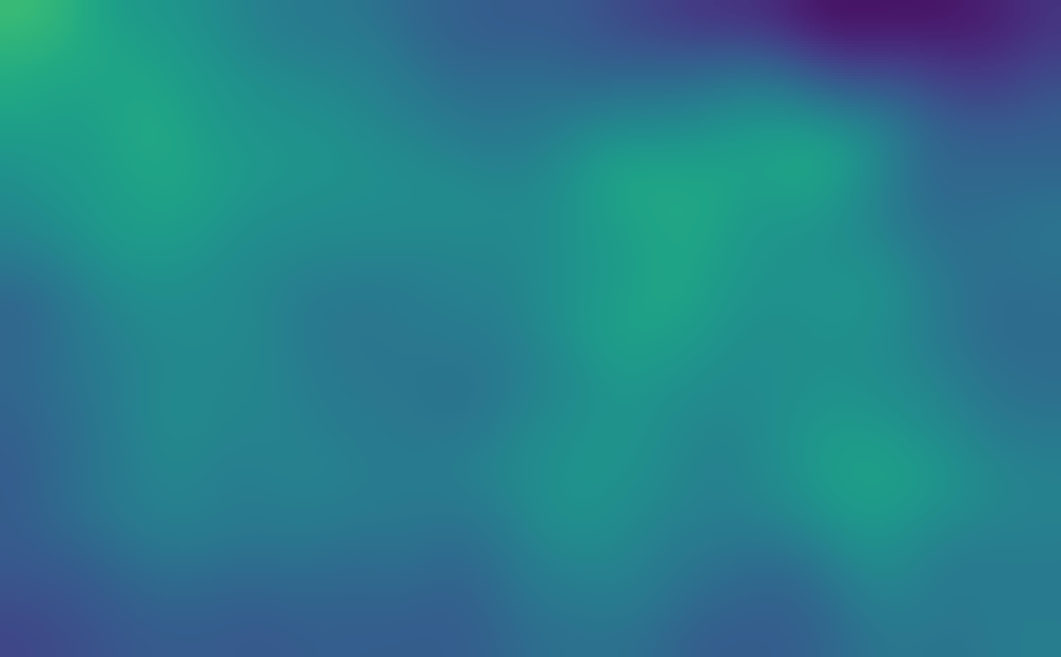} \\
        \multirow{2}{*}{\rotatebox{90}{\shortstack{U-Net \\ (multi)}}} 
            & \includegraphics[width=\imwidth\linewidth]{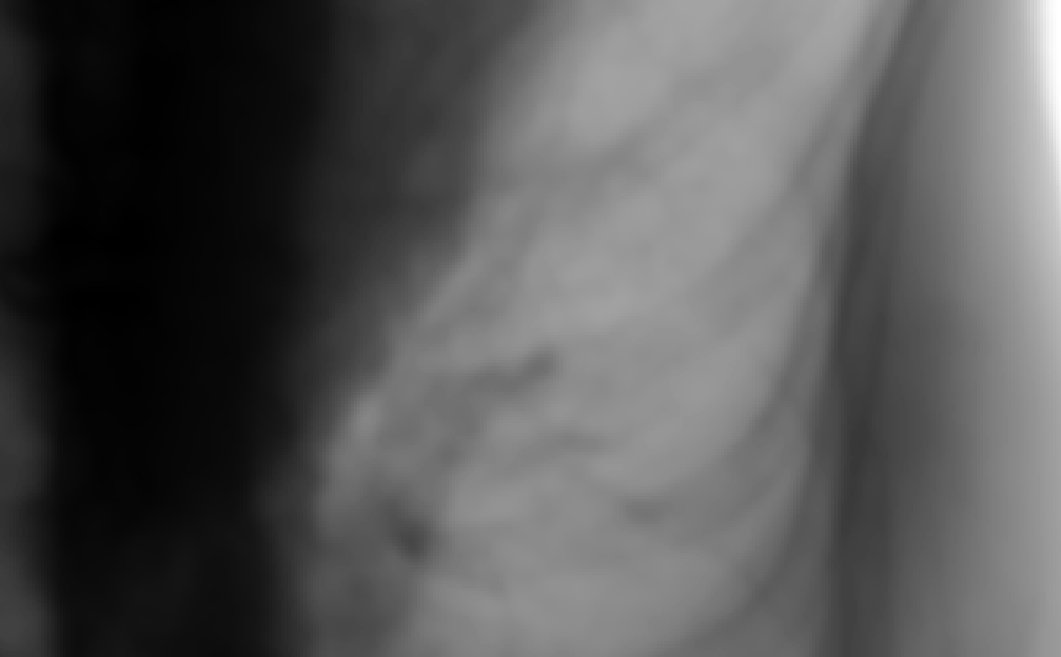}
            & \includegraphics[width=\imwidth\linewidth]{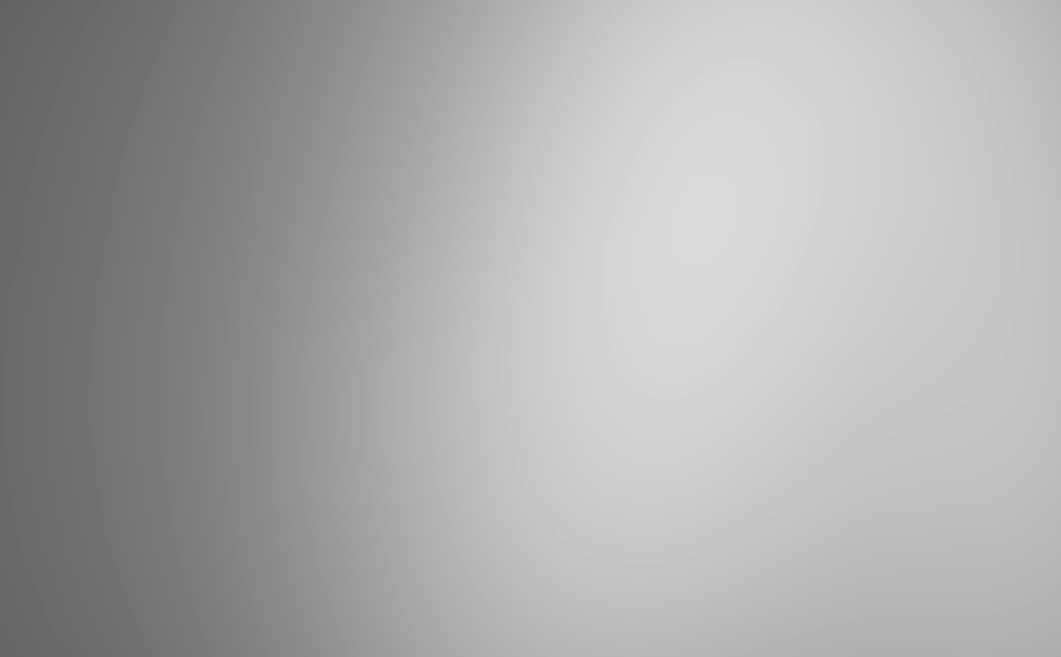}
            & \includegraphics[width=\imwidth\linewidth]{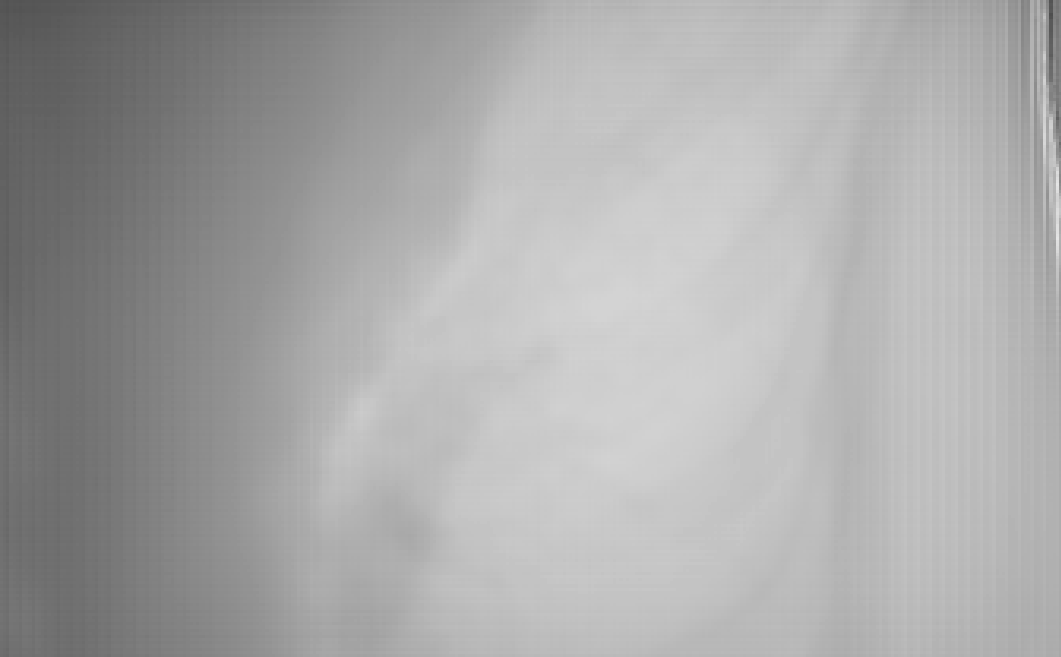}
            & \includegraphics[width=\imwidth\linewidth]{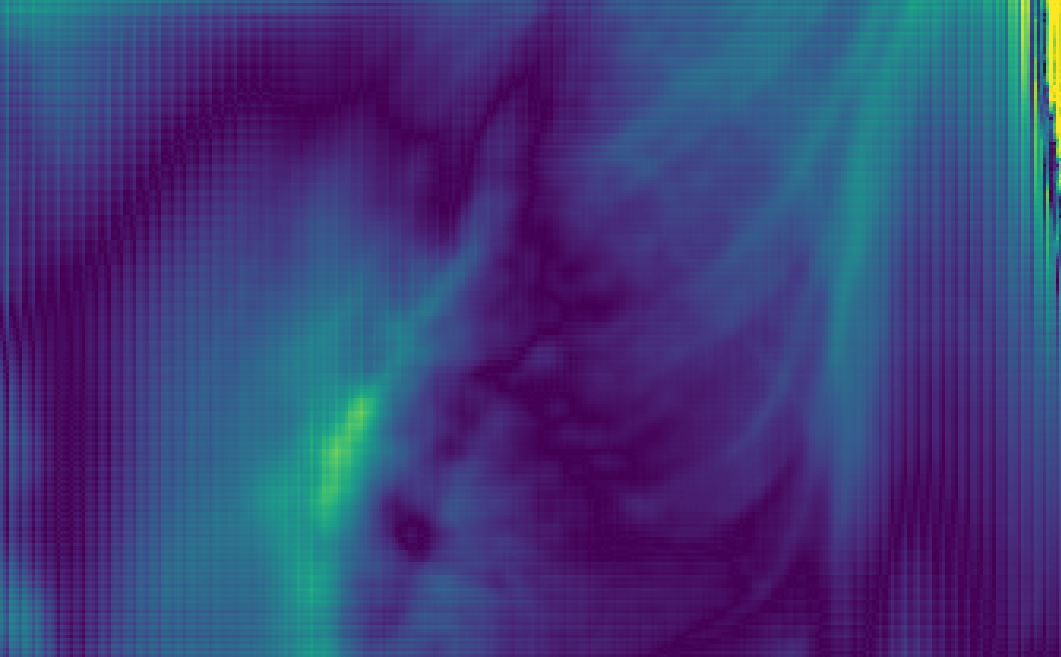}
            & \includegraphics[width=\imwidth\linewidth]{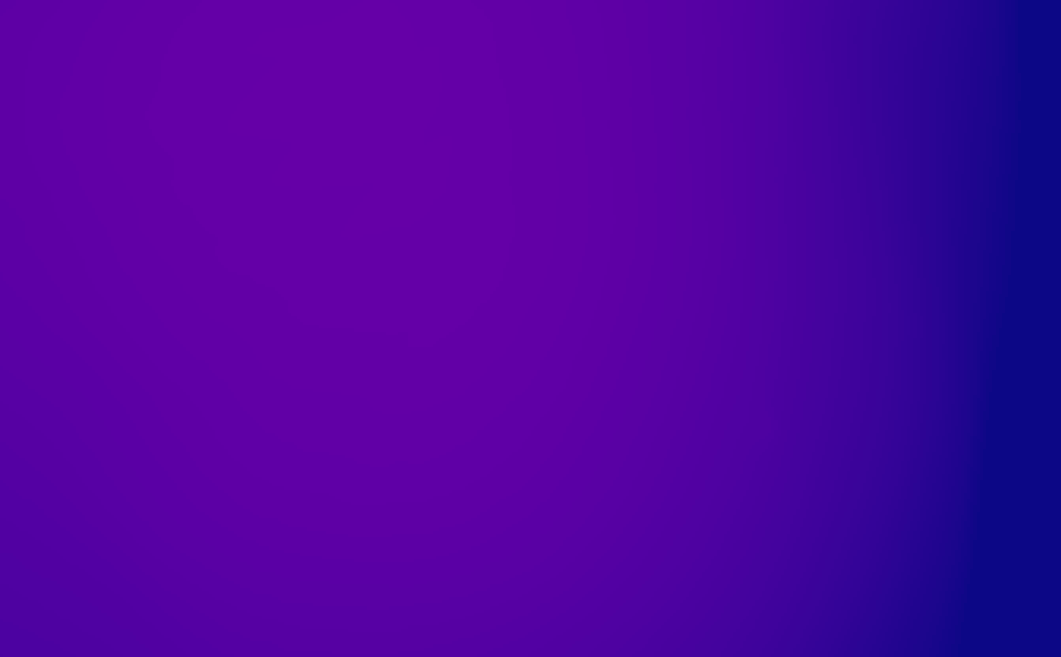}
            & \includegraphics[width=\imwidth\linewidth]{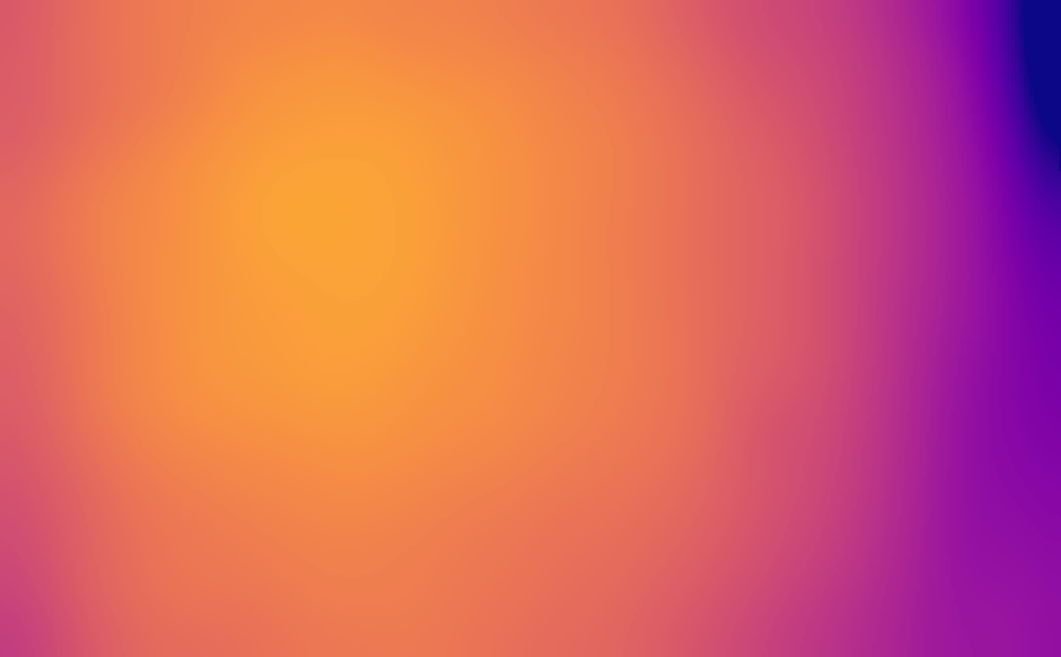}
            & \includegraphics[width=\imwidth\linewidth]{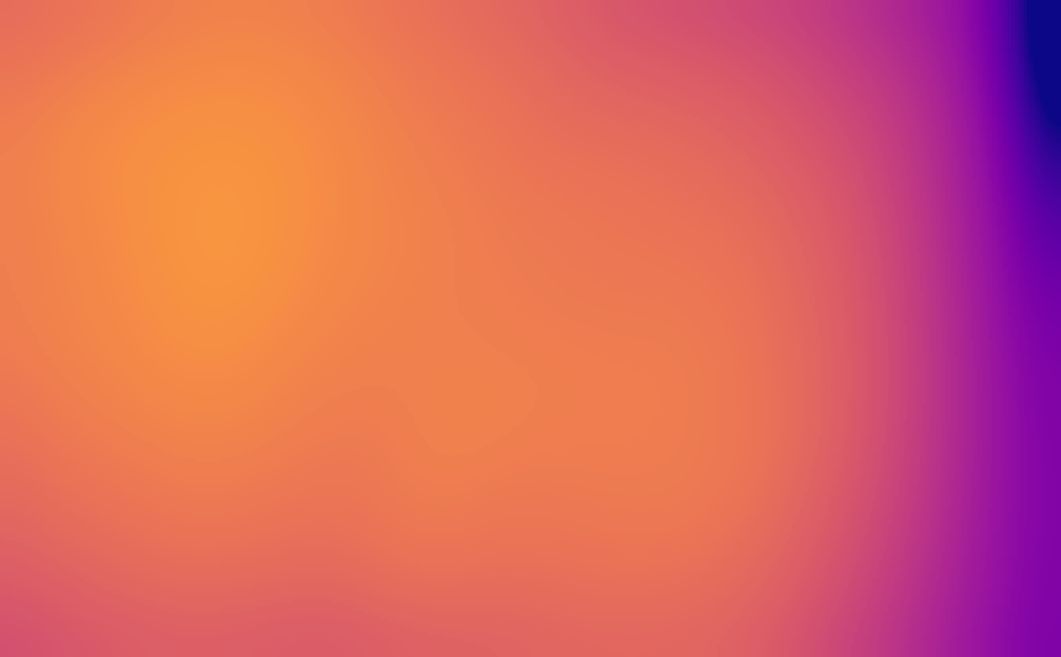}
            & \includegraphics[width=\imwidth\linewidth]{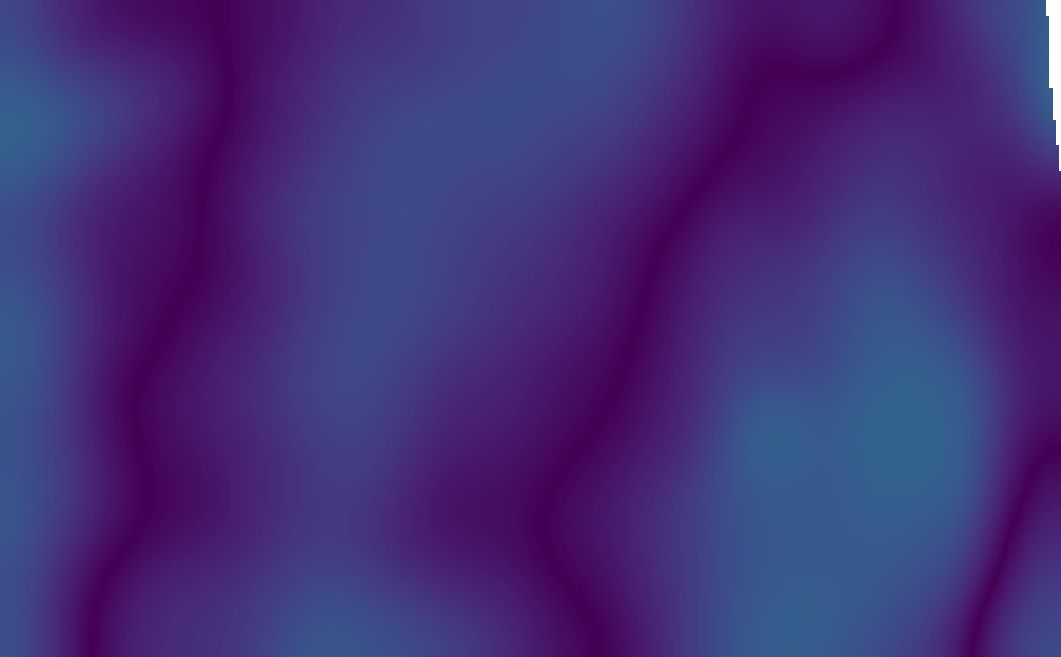} \\
            & \includegraphics[width=\imwidth\linewidth]{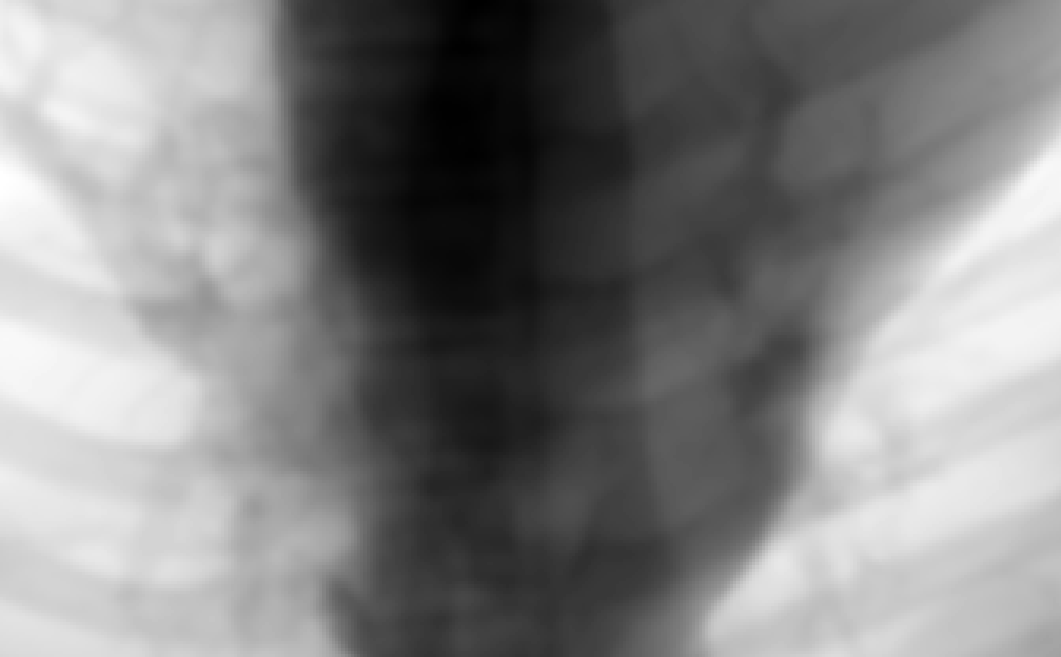}
            & \includegraphics[width=\imwidth\linewidth]{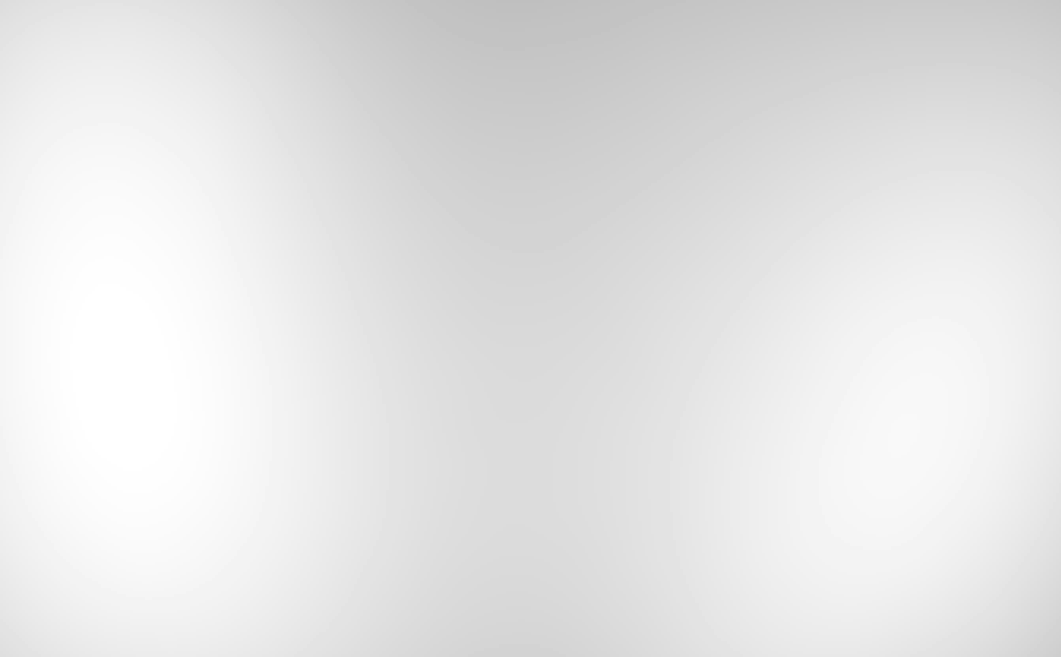}
            & \includegraphics[width=\imwidth\linewidth]{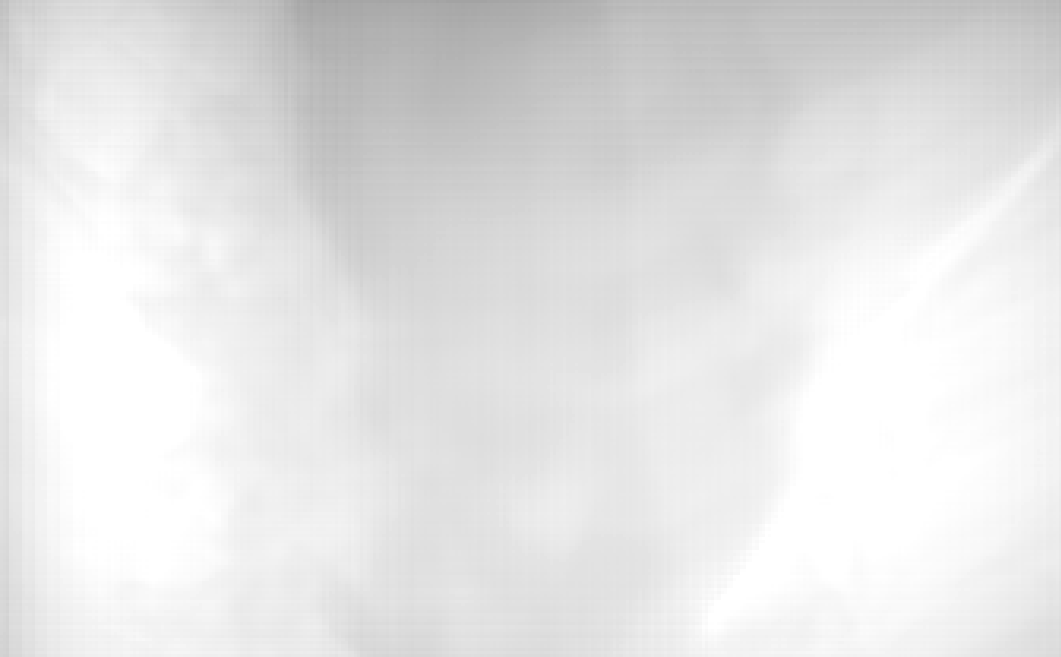}
            & \includegraphics[width=\imwidth\linewidth]{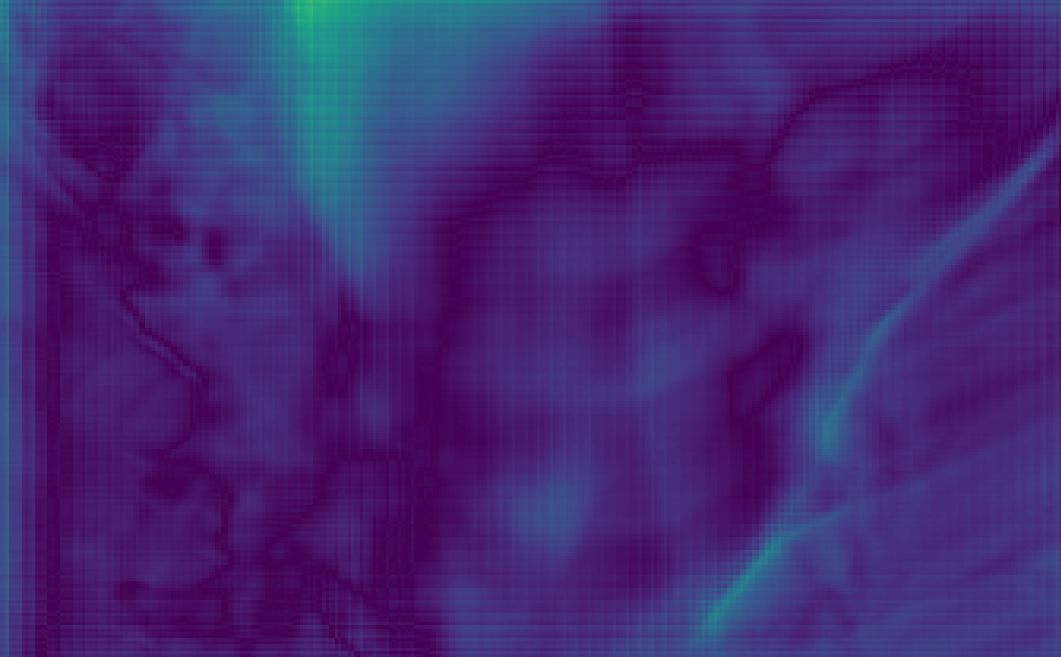}
            & \includegraphics[width=\imwidth\linewidth]{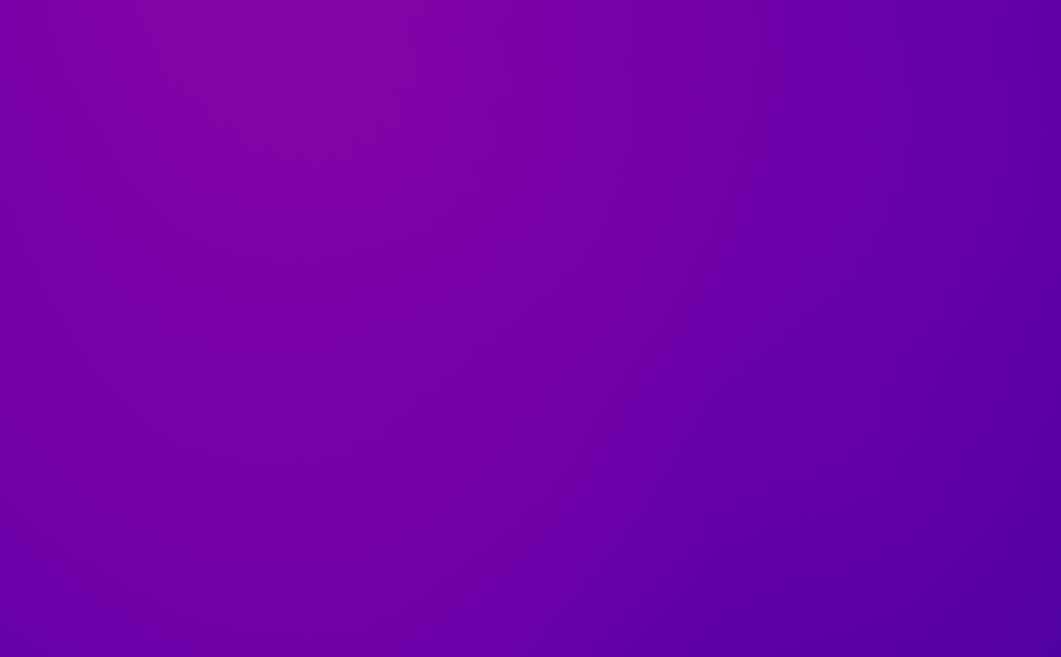}
            & \includegraphics[width=\imwidth\linewidth]{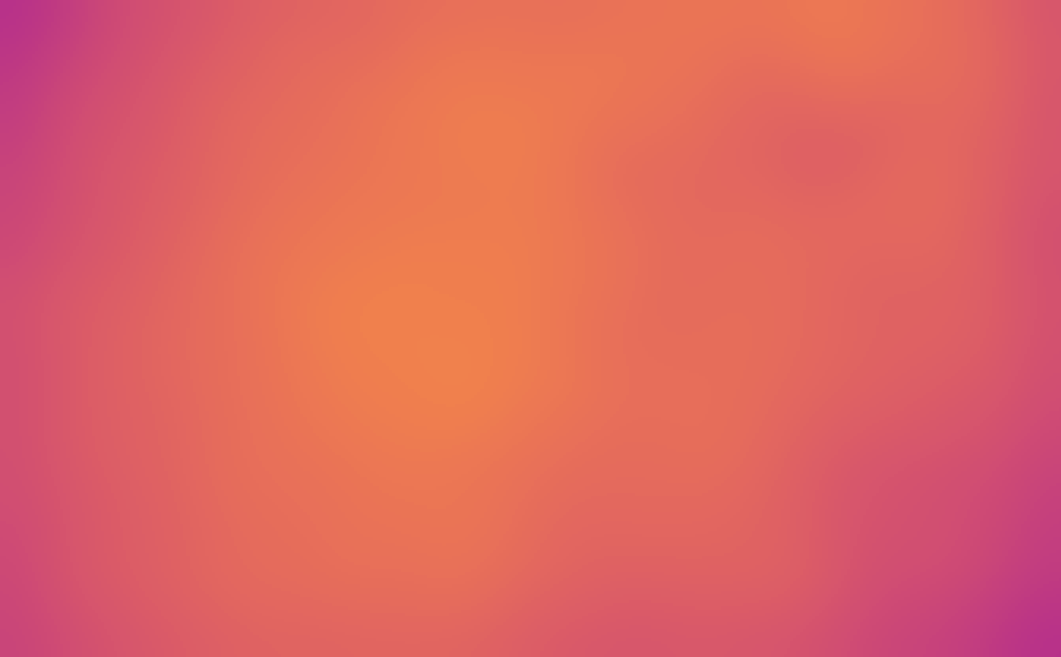}
            & \includegraphics[width=\imwidth\linewidth]{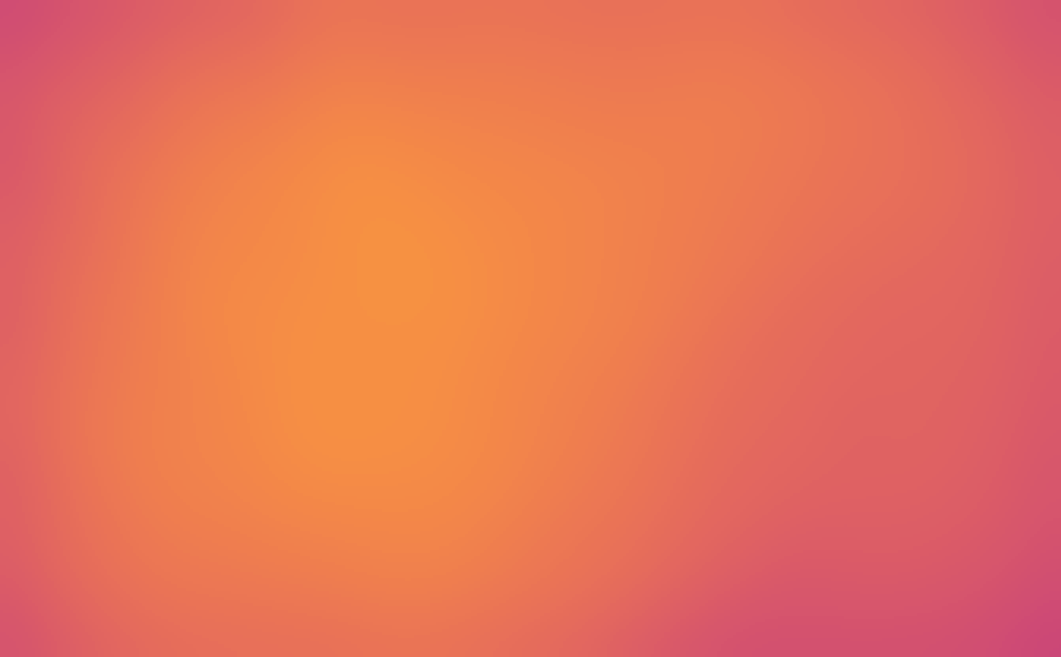}
            & \includegraphics[width=\imwidth\linewidth]{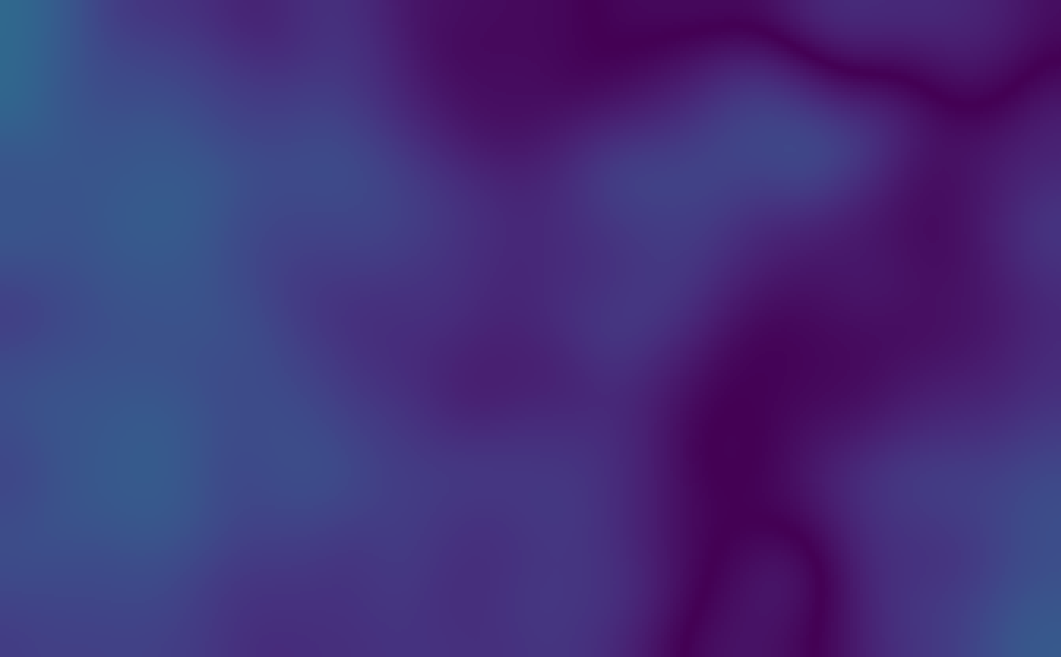} \\
        \multirow{2}{*}{\rotatebox{90}{DAE}} 
            & \includegraphics[width=\imwidth\linewidth]{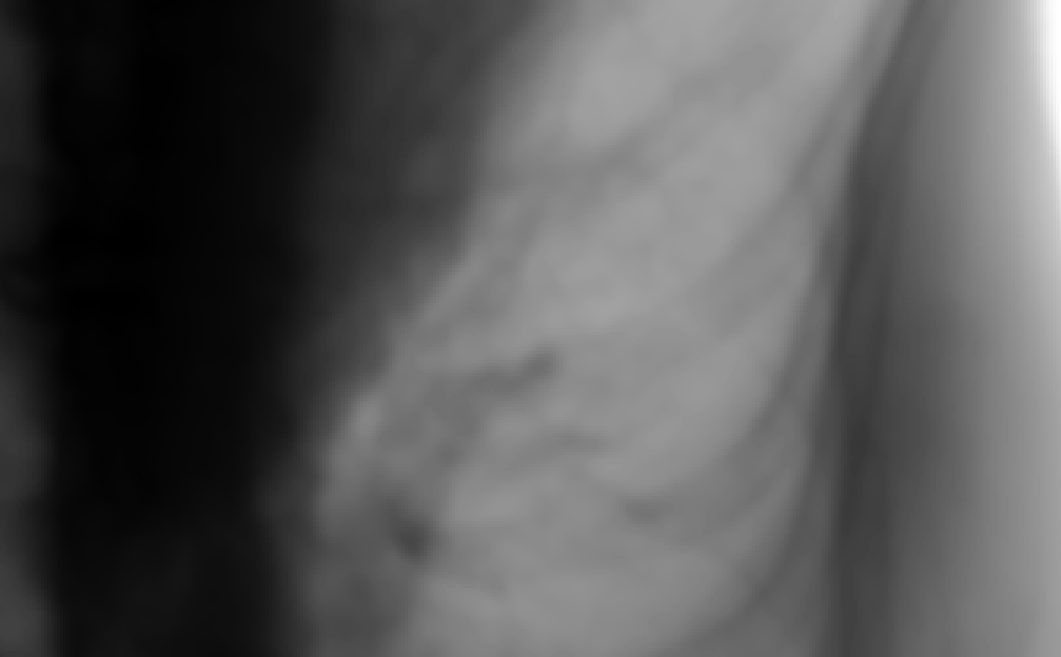}
            & \includegraphics[width=\imwidth\linewidth]{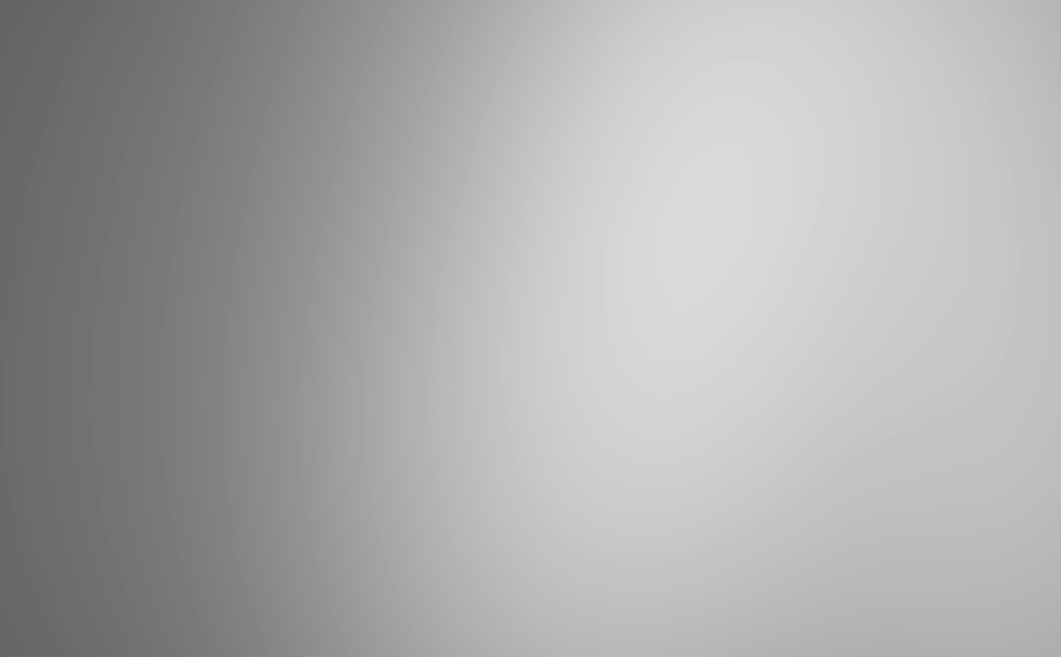}
            & \includegraphics[width=\imwidth\linewidth]{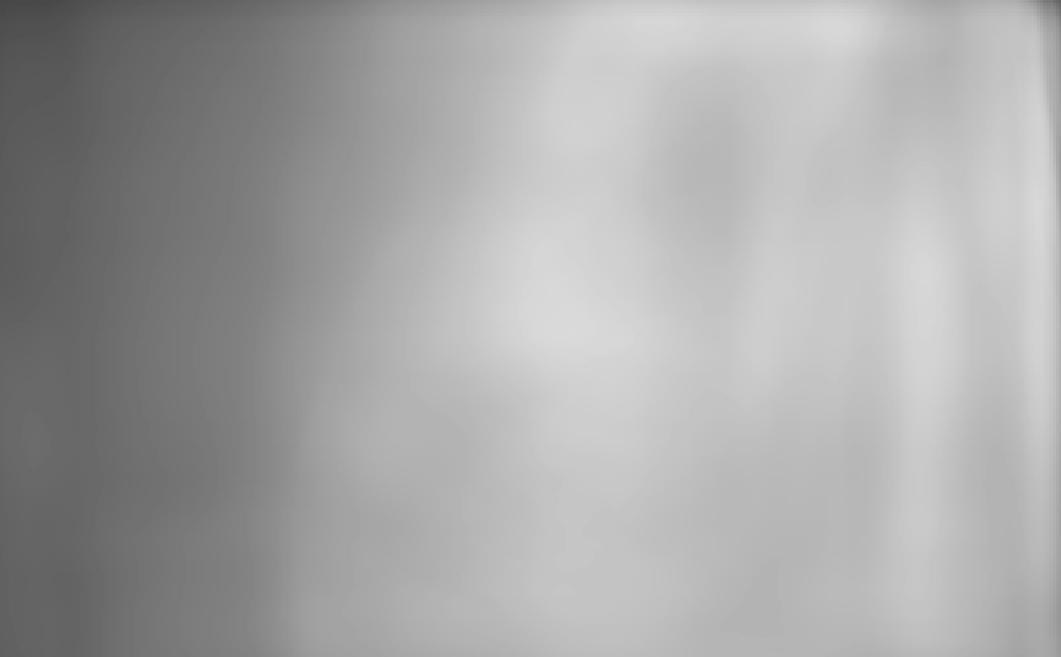}
            & \includegraphics[width=\imwidth\linewidth]{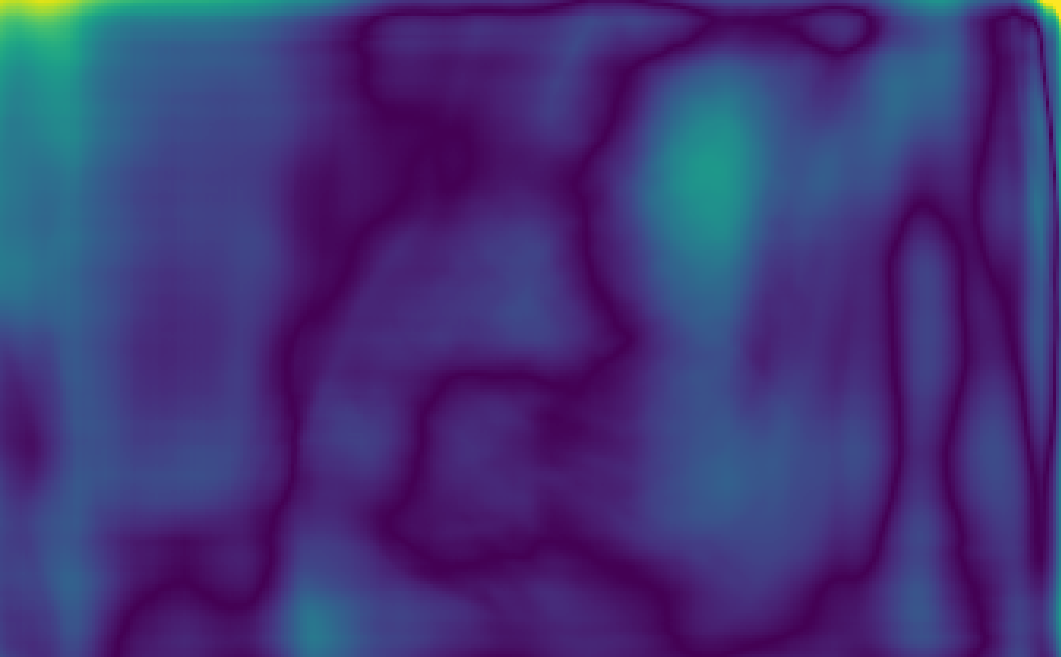}
            & \includegraphics[width=\imwidth\linewidth]{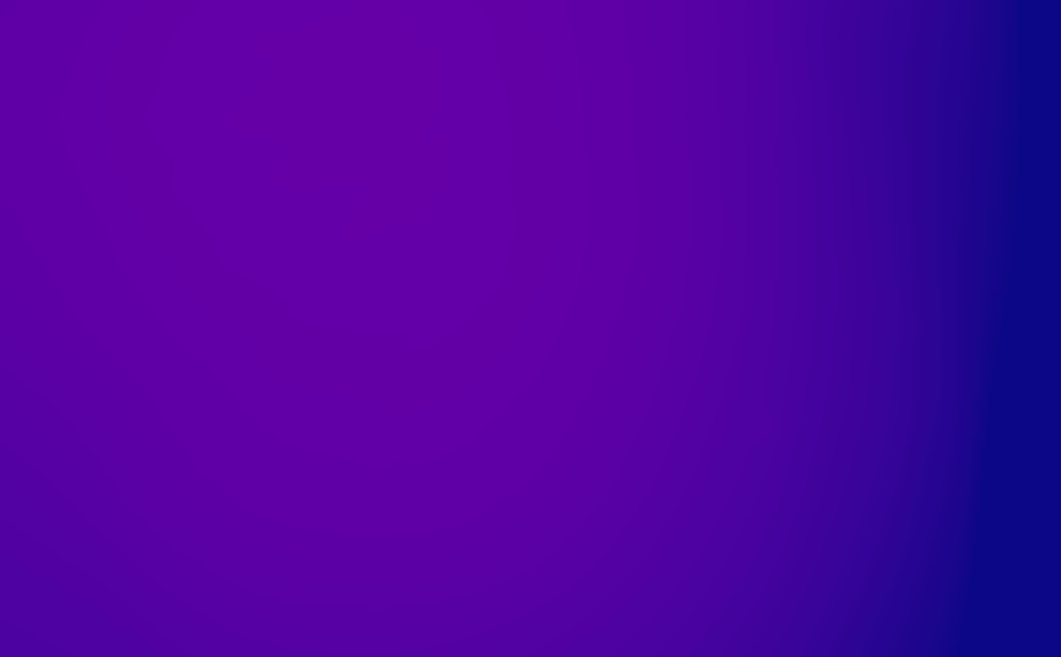}
            & \includegraphics[width=\imwidth\linewidth]{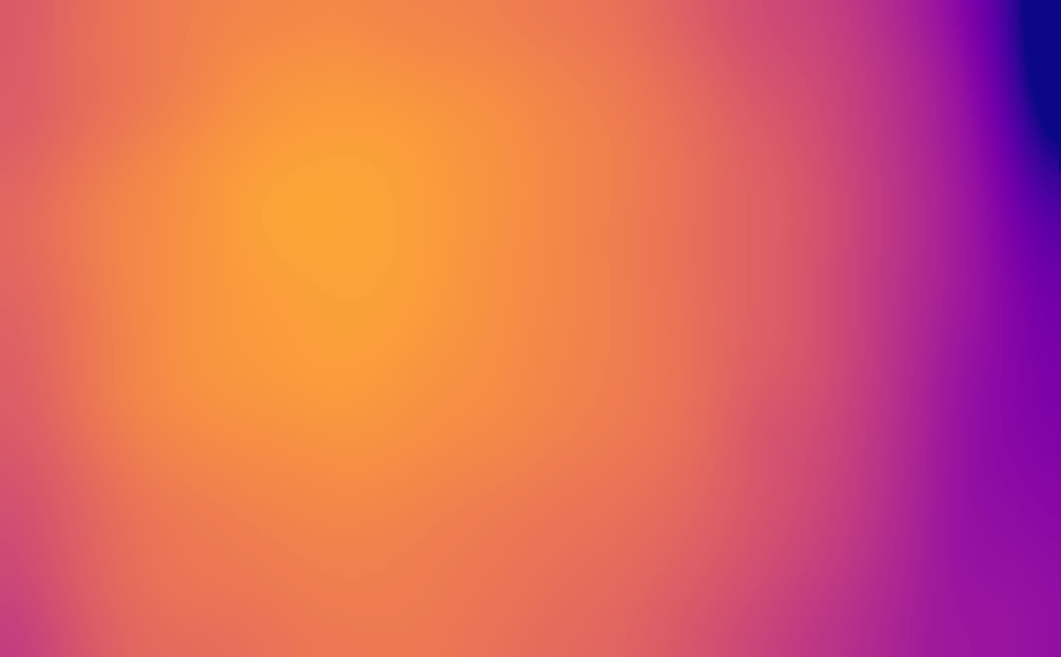}
            & \includegraphics[width=\imwidth\linewidth]{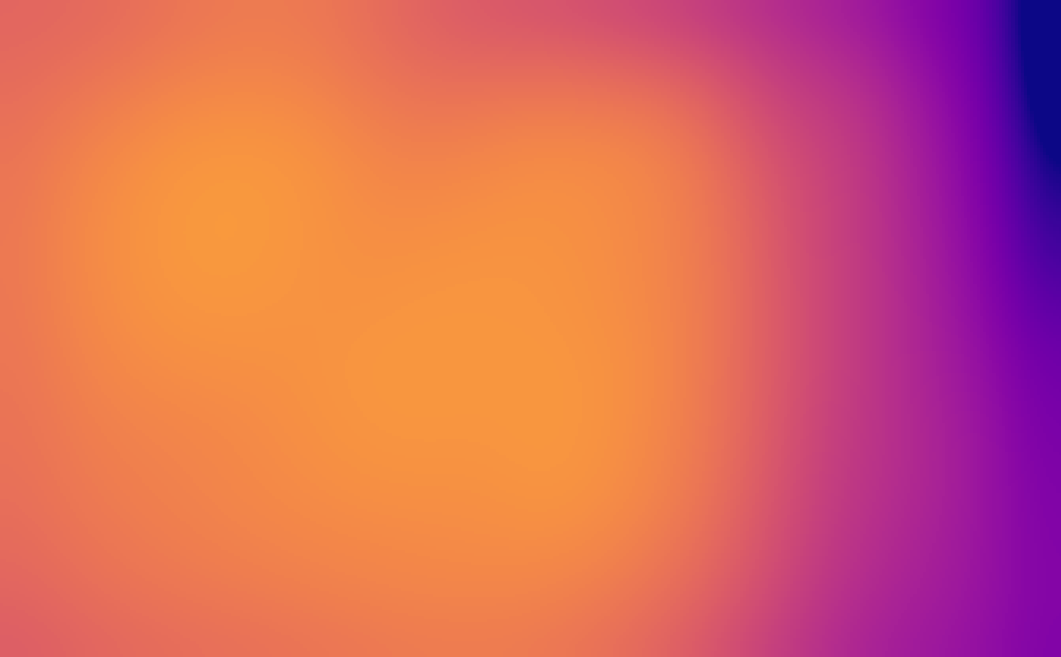}
            & \includegraphics[width=\imwidth\linewidth]{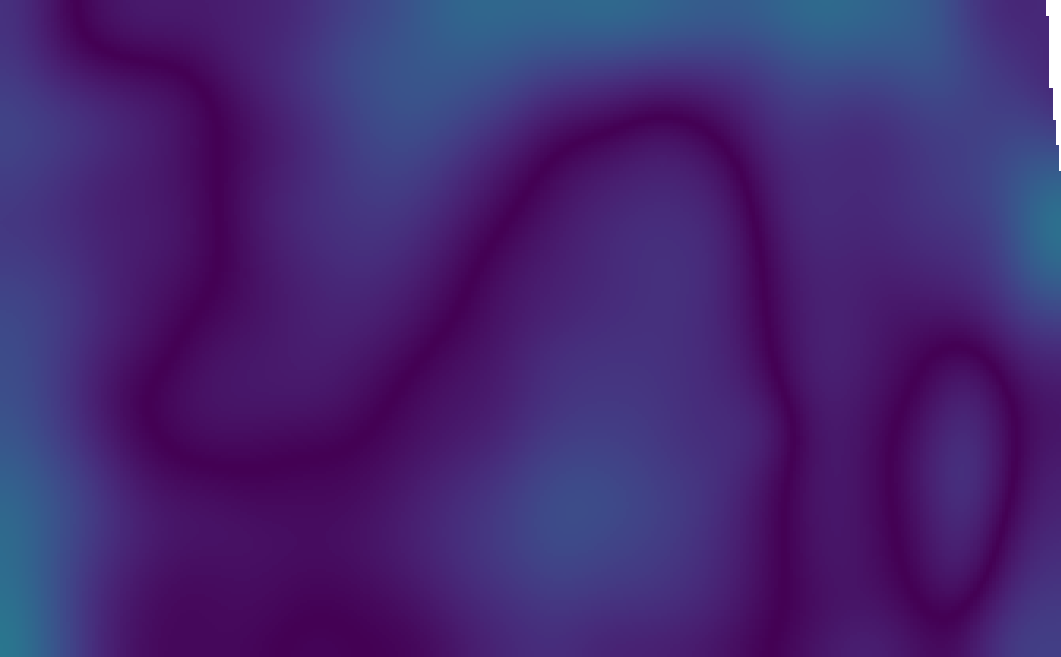} \\
            & \includegraphics[width=\imwidth\linewidth]{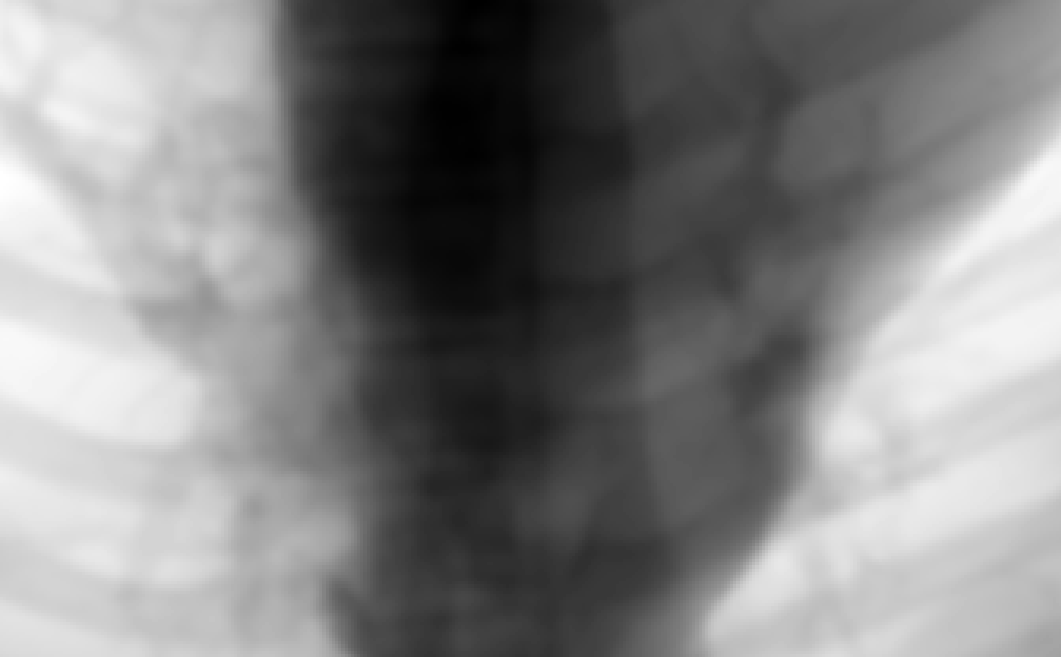}
            & \includegraphics[width=\imwidth\linewidth]{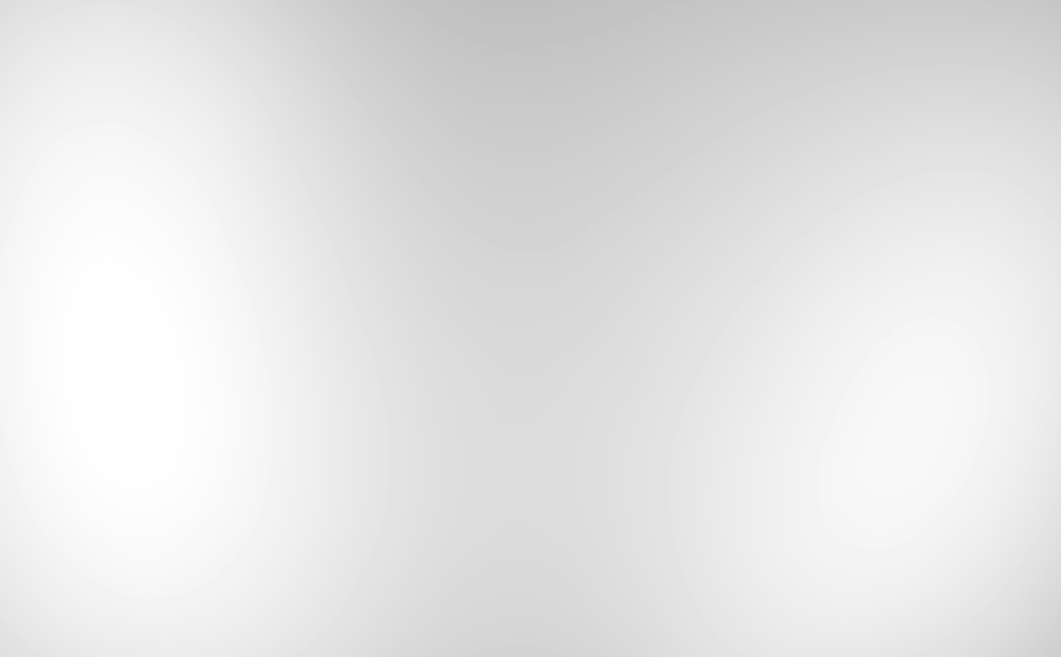}
            & \includegraphics[width=\imwidth\linewidth]{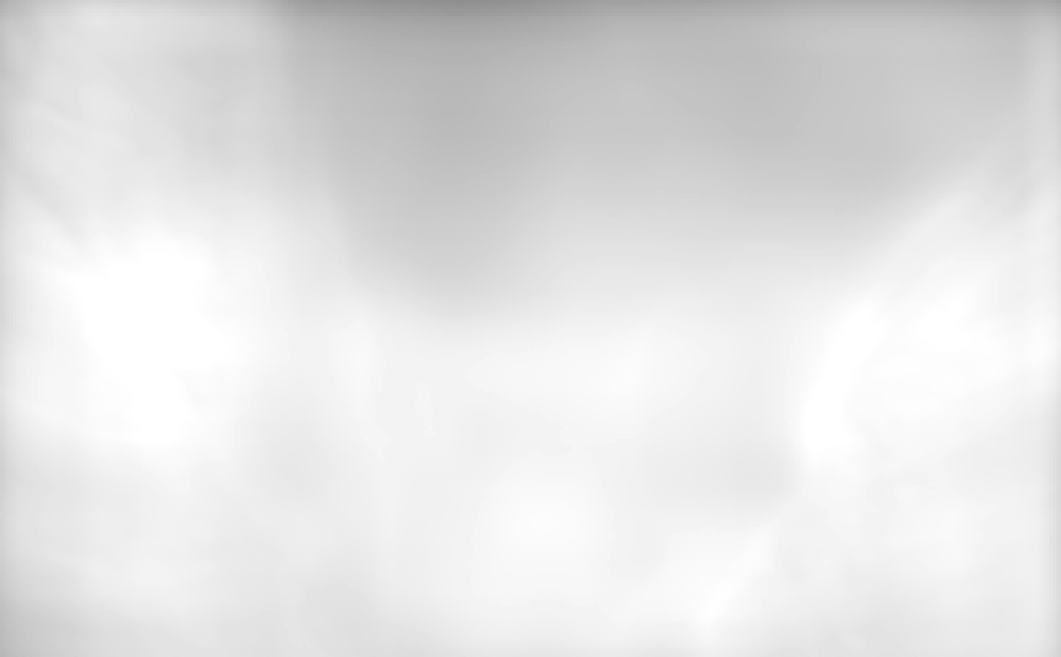}
            & \includegraphics[width=\imwidth\linewidth]{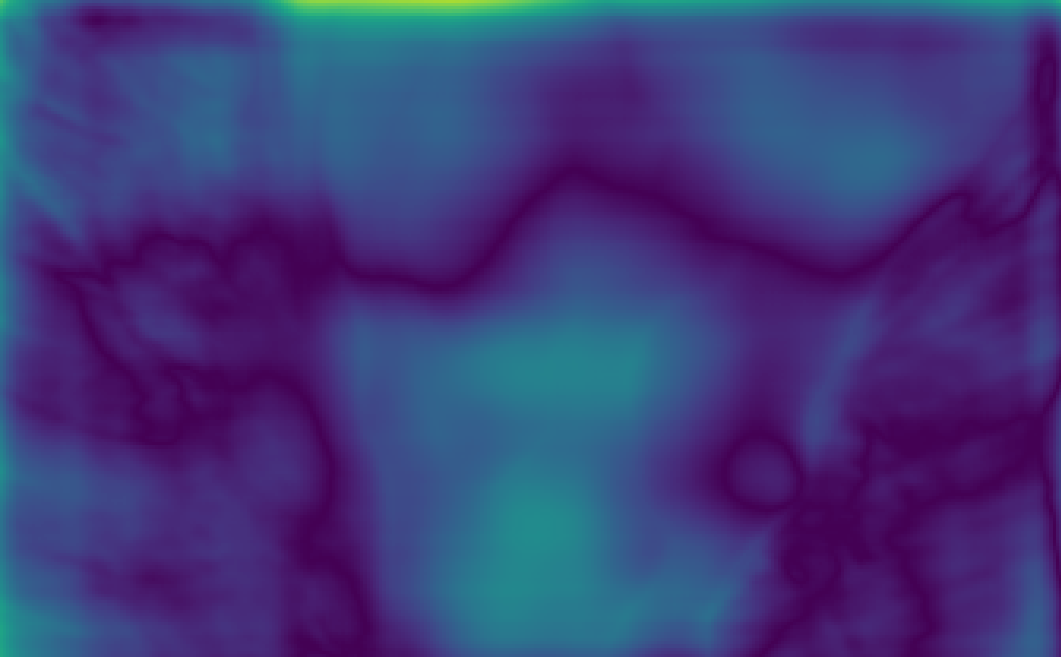}
            & \includegraphics[width=\imwidth\linewidth]{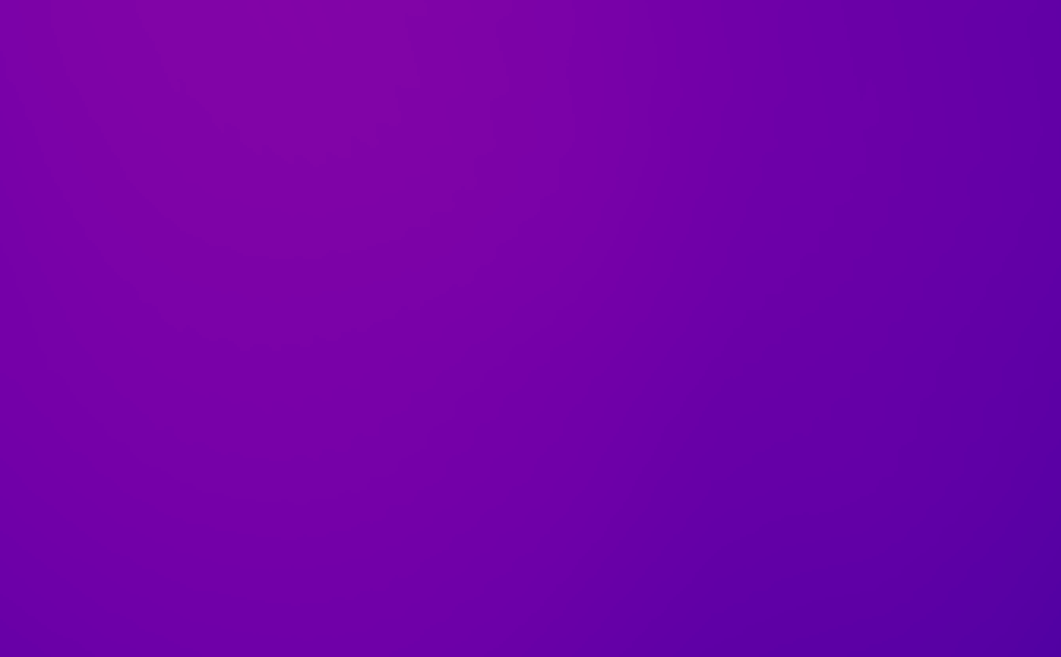}
            & \includegraphics[width=\imwidth\linewidth]{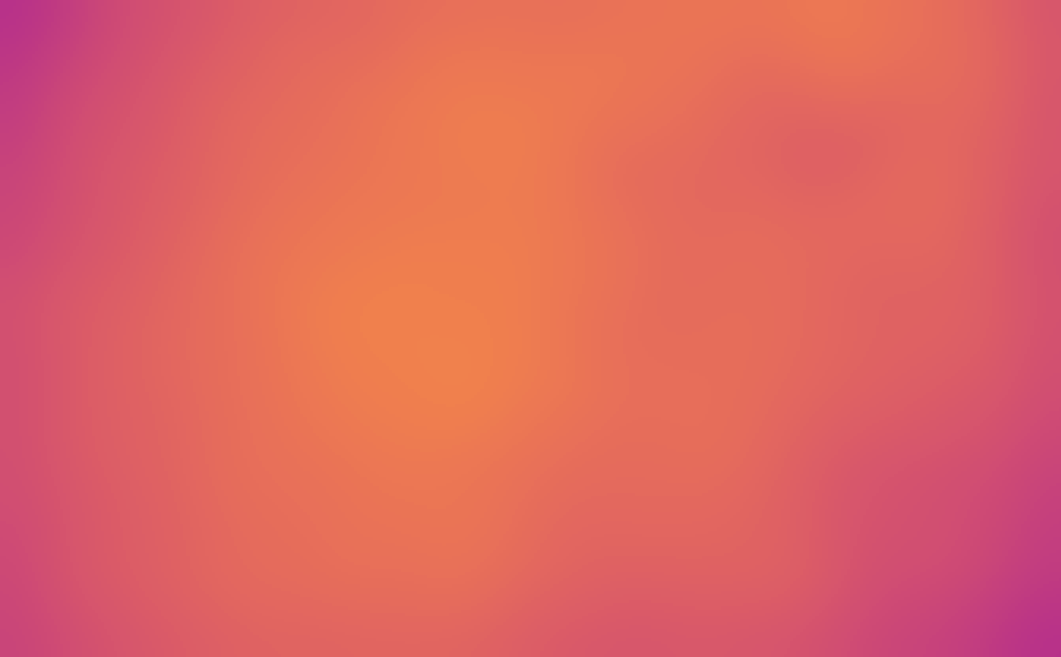}
            & \includegraphics[width=\imwidth\linewidth]{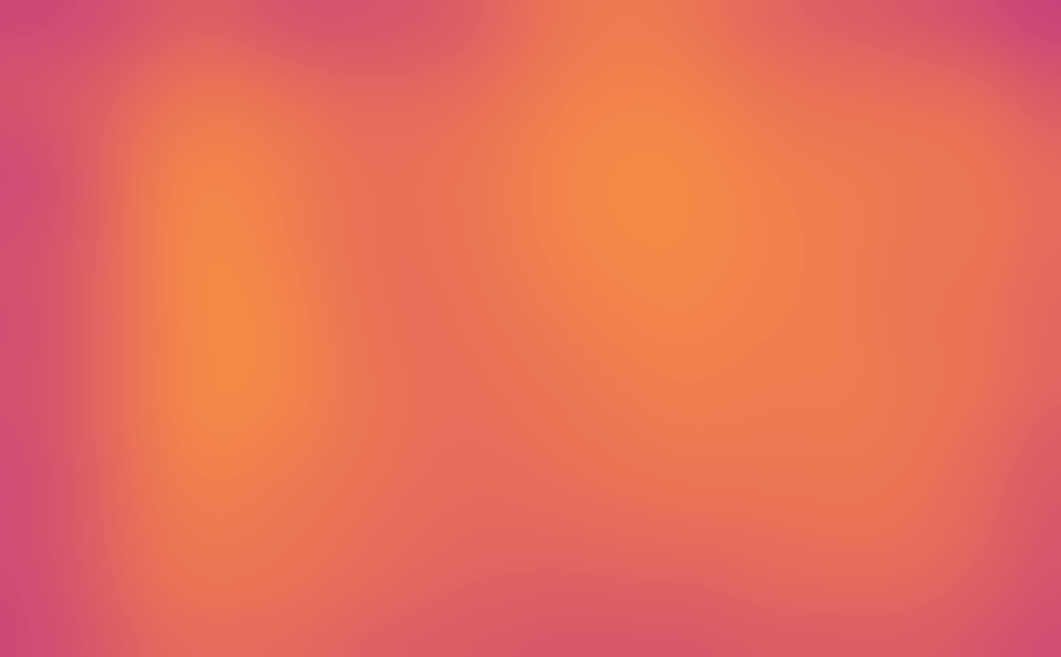}
            & \includegraphics[width=\imwidth\linewidth]{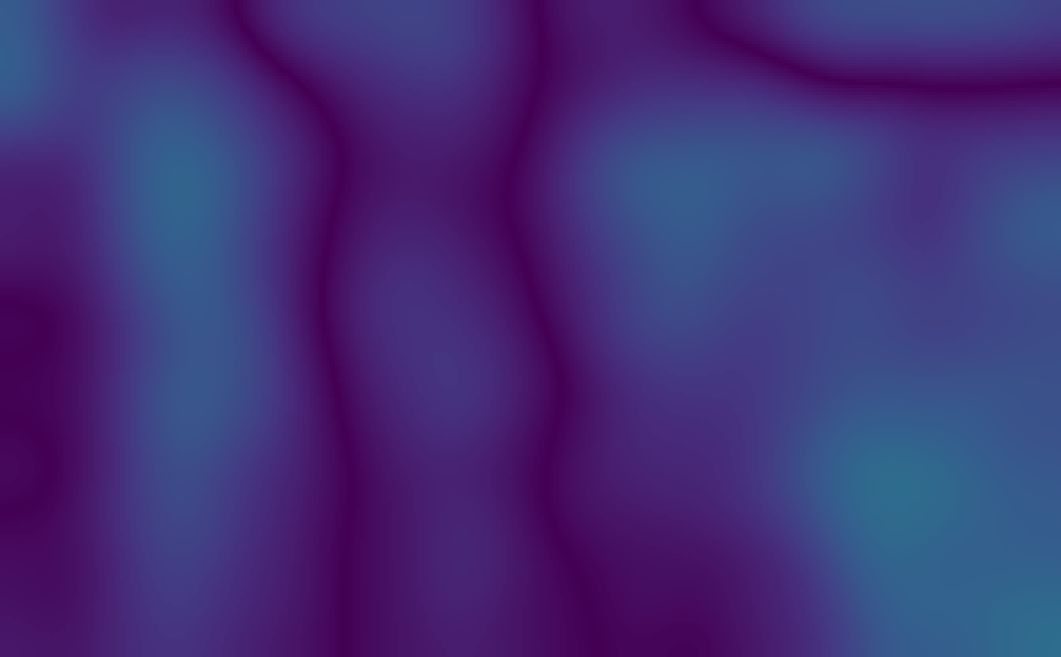} \\
    \end{tabular}
    \caption{
        Two qualitative examples of both U-Nets and the DAE for the thorax data set. 
        From left to right: The measured X-ray projection $\vec{I}_\text{t}$, corresponding forward-scatter ground truth $\vec{I}_\text{s}$ with associated network estimates $\hat{\vec{I}}_\text{s}$ and relative error maps $\vec{\varepsilon}_{\vec{I}}$, and the primary skin dose $\vec{D}_\text{p}$, corresponding back-scatter ground truth $\vec{D}_\text{s}$ with associated network estimates $\hat{\vec{D}}_\text{s}$ and relative error maps $\vec{\varepsilon}_{\vec{D}}$. 
        Error maps are in the range of \SIrange{0}{30}{\percent} (dark blue over green to yellow).
    }
    \label{fig:results}
\end{figure}
Table \ref{tab:results} summarizes expected error rates for all network settings for the testing data.
For the head data set, both, the single-task AE and the single-task U-Net extracted $\vec{I}_\text{s}$ from X-ray projections with similar and high accuracy.
Nevertheless, both single-task networks leave room for improvement concerning skin dose estimation.
Overall the multi-task U-Net and DAE approaches clearly outperformed the single-task ones, especially in terms of back-scatter dose estimation.
However, the multi-task AE approach appears to lack the capacity to estimate $\vec{D}_\text{s}$ and $\vec{I}_\text{s}$ simultaneously.
Instead it focused on one quantity with high accuracy (see head data) or both with lower accuracy (see thorax data).
Overall, the multi-task U-Net found the best mapping for all test cases.
Due to its high number of parameters, the multi-task U-Net can map the relationship between both scatter types easily.
With our modifications of the AE, the resulting DAE network, however, performed almost on par with the multi-task U-Net, but it achieved this with a at much lower parameter complexity.
Figure \ref{fig:results} combines exemplary qualitative results of both U-Nets and the DAE.
The depicted error maps substantiate the quantitative results we observed.
They also reveal current limitations of our approach.
Especially the U-Net preserved spurious structural information in the scatter distributions due to its high number of parameters to train, and skip connections.
Although the DAE performed worse in terms of quantitative results, it did not hallucinate high-frequency details (edges) to the same degree.
Besides, with an average runtime of \SI{100}{\milli\second}, it was more than twice as fast as the U-Net with a runtime of \SI{240}{\milli\second} (CPU, Intel(R) i7-8850H), suggesting that it can be used in a clinical setup once current limitations are solved.
Promising counter-measures include the incorporation of more prior knowledge, such as deriving scatter probabilities based on the patient shape model, or the combination with a first-order scatter estimation algorithm \cite{Freud:2004:FirstOrder,Ingleby:2015:FirstOrder,Yao:2009:FirstOrder}.

\section{Conclusion}
We presented an multi-task learning-based framework to (a) estimate the back-scatter contribution to the total skin dose and (b) estimate the forward-scatter in the X-ray projection image.
To the best of our knowledge, this is the first paper investigating back-scatter estimation in an learning-based fashion and to then combine it with forward-scatter calculation.
For the AE and the U-Net, we showed that, by estimating back- and forward-scatter in an multi-task fashion, the accuracy in both cases benefits compared to their respective single-task versions.
We identified limitations in our approach and proposed appropriate counter-measures for future work.
In addition, with only minor adjustments of the AE architecture, we achieved almost the same accuracy as the multi-task U-Net, while decreasing the parameter complexity and thus increasing the computational efficiency more than twofold.
Especially in an interventional environment, where scatter is a critical aspect, our approach has the potential to facilitate dose reduction while maintaining or even improving image quality.

~\\
\textbf{Disclaimer~} The concepts and information presented in this paper are based on research and are not commercially available.


\end{document}